\documentclass[twocolumn]{aastex631}

\begin{document}

\title{Multi-wavelength observations of a B-class flare using XSM, AIA, and XRT}

\author{Yamini K. Rao}
\affiliation{DAMTP, University of Cambridge, Wilberforce Rd, Cambridge CB3 0WA, United Kingdom }

\author{B. Mondal}
\affiliation{Physical Research Laboratory, Navrangpura, Ahmedabad, Gujarat-380 009, India}

\author{Giulio Del Zanna}
\affiliation{DAMTP, University of Cambridge, Wilberforce Rd, Cambridge CB3 0WA, United Kingdom }

\author{N. P. S. Mithun}
\affiliation{Physical Research Laboratory, Navrangpura, Ahmedabad, Gujarat-380 009, India}

\author{S. V. Vadawale}
\affiliation{Physical Research Laboratory, Navrangpura, Ahmedabad, Gujarat-380 009, India}

\author{K. K. Reeves}
\affiliation{Center for Astrophysics | Harvard \& Smithsonian, 60 Garden St. MS 58. Cambridge, MA 02138}

\author{Helen E. Mason}
\affiliation{DAMTP, University of Cambridge, Wilberforce Rd, Cambridge CB3 0WA, United Kingdom }

\author{Anil Bhardwaj}
\affiliation{Physical Research Laboratory, Navrangpura, Ahmedabad, Gujarat-380 009, India}

\begin{abstract}
We present multi-wavelength observations by Chandrayaan-2/XSM, SDO/AIA and Hinode/XRT of a B-class flare observed on 25th February, 2021, originating from an active region (AR 12804) near the North-West limb. The microflare lasts for $\sim$ 30 mins 
and is composed of hot loops reaching temperatures of 10 MK. We report excellent agreement (within 20\%) for the average effective temperatures obtained at the flare peak from all the three instruments, which have different temperature sensitivities. The XRT filter combination of Be-thin and Be-med provides an excellent opportunity to measure the  high-temperatures in such microflare events. The elemental abundances during the evolution of the microflare are also studied and observed to drop towards photospheric values at the flare peak time, compared to coronal values during the rise and decay phase. This is consistent with previous XSM studies.
\end{abstract}

\section{Introduction} 
\label{sec:intro}
Small solar flares frequently occur in active regions.
We are particularly interested in microflares, which we loosely define as events of GOES X-ray class B and lower.
During microflares, small loop structures of plasma having temperatures above the  background in Active Region (AR) cores (typically around 2$-$3 MK) become activated. They typically have short lifetimes (ten minutes or so).
 These small-scale energy releases offer insights into how energy and mass are transferred from the lower atmosphere to the solar corona, and are simpler to study than the larger events. 
 In spite of numerous studies of larger flares (M or X$-$class) in order to understand solar flare heating, researchers remain interested in investigating why flares of the same class can have vastly different peak temperatures. Microflares can be studied through a combination of observations in the EUV and X$-$ray wavelength domains, with X$-$ray emission being especially sensitive to higher temperatures. 

\begin{center}
\begin{figure*}
\includegraphics[width=1\linewidth,angle=180]{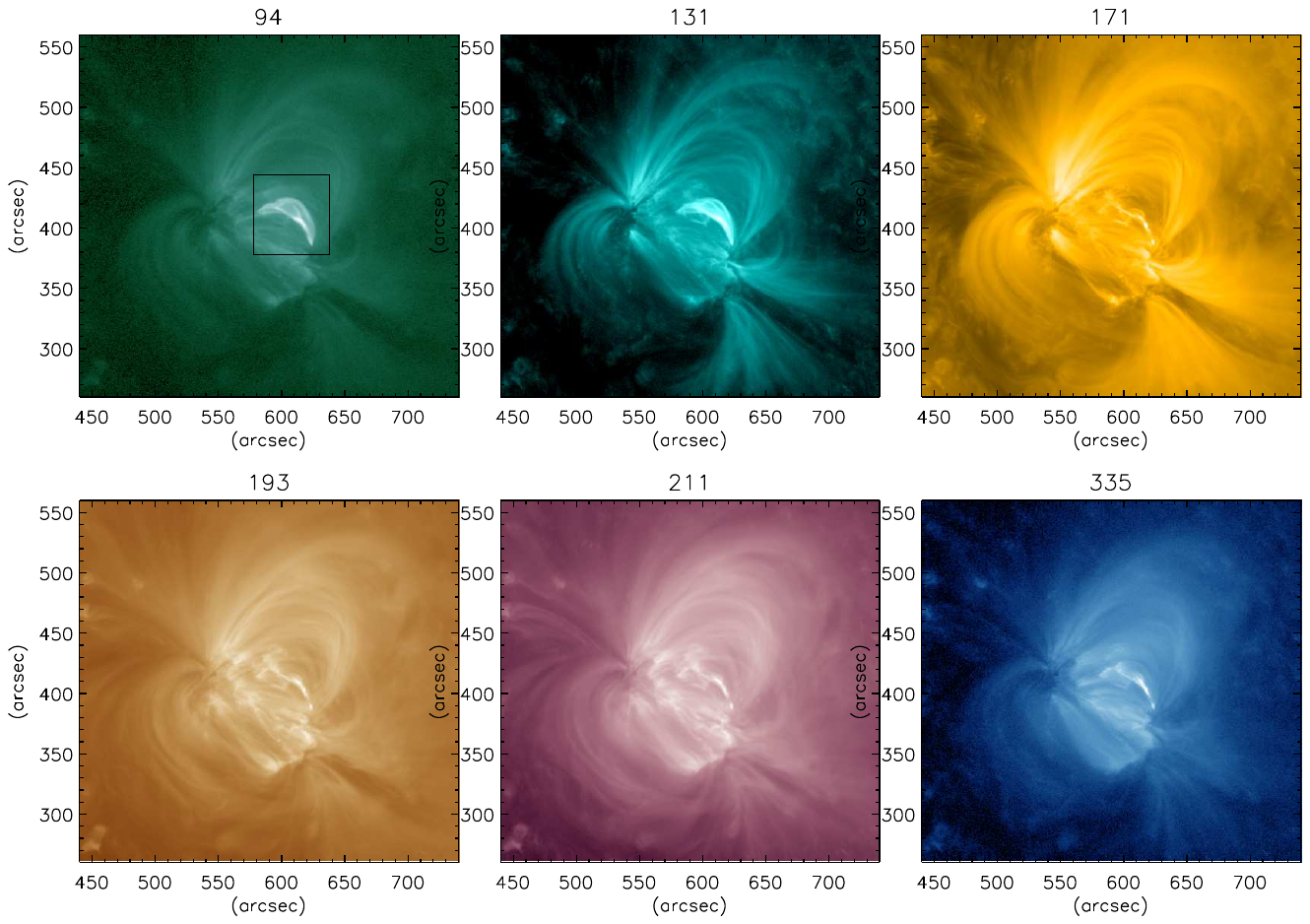}
\caption{The SDO/AIA images observed in multi-wavelength channels targeting the AR 12804 observed on 25th February 2021 at flare peak time (14:16 UT). The wavelengths of different channels are noted at the top of the respective panels. The black box on the 94 \AA\ image shows the region of the flare loops, shown in more detail in the following figures.}
\label{fig:aia}
\end{figure*}
\end{center}

The Solar X-ray Monitor (XSM) onboard Chandrayaan-2 provides disk-integrated spectra in the energy range of 1-15 keV. XSM has a high sensitivity and can detect small as well as large solar activity from sub-A to X-class, with a good signal-to-noise ratio. Modelling the XSM spectrum can provide an accurate measurement of the plasma temperature, emission measure, and the abundances of several elements (e.g., Mg, Al, Si, S, Fe etc.).
Using the XSM observations during the minimum of solar cycle 24, \cite{xsm_microflares_2021} reported the detection of a large number of sub-A class flares outside of ARs.
These observations were also used to measure the plasma parameters of the global quiet-Sun plasma with an average temperature $\sim$2 MK, which was dominated by the X-ray bright points (XBPs:~\citealp{xsm_XBP_abundance_2021,2023ApJ...945...37M}).

\begin{center}
\begin{figure*}
\includegraphics[width=1\linewidth,angle=180]{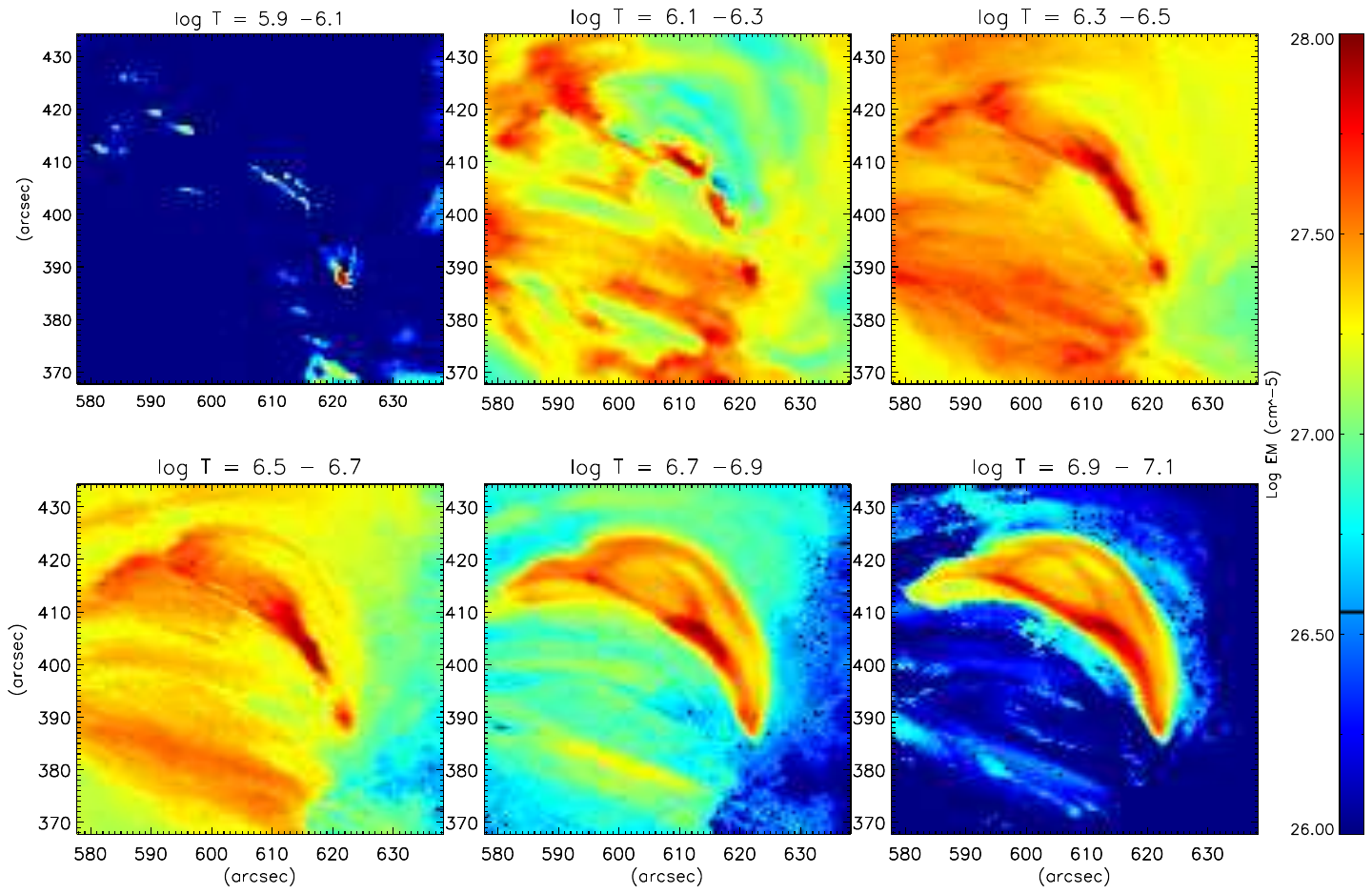}
\caption{Emission measure maps of the AR data at the flare peak time (averaged for a 3-min interval, 14:14 to 14:17 UT).}
\label{fig:dem}
\end{figure*}
\end{center}

XSM also observed multiple flaring in ARs and these data have been used to study the time evolution of the plasma parameters during A-class, B-class and C-class flares~\citep{laksitha_2022, Mondal_2021ApJ,Mithun_2022}.
These studies show that during the impulsive phase, the temperature of the flaring plasma reaches a few MK, whereas elemental abundances show a sharp transition from coronal to close to photospheric values. The abundances were found to return to coronal values during the decay phase of the flares. 
Such variations provide new observational constraints for flare modelling.
To complement the XSM observations, \cite{Zanna_2022ApJ} carried out a multi-wavelength campaign of an AR (NOAA 12759) that was observed by XSM as well as SDO/AIA, XRT and EIS onboard Hinode. They found discrepancies between the filter-ratio temperatures from XRT and 
the values measured by the other instruments. Consistency in the observations was found with emission measure modelling. 
The XRT filter ratio used for the campaign, Al-Poly vs. Be-thin, is sensitive to a broad range of temperatures, providing values significantly lower than those measured by XSM.

Based on these findings, we modified the XRT Observing sequence HOP (Hinode Observing Programme) 396 to use the Be-med vs Be-thin filter for further coordinated observations.
One of the outstanding question about microflares is the relationship between the peak temperatures  and the X-ray class. 
\cite{2011A&A...526A...1D} found temperatures as high as 12 MK in a B2-class microflare, while the B-9 class microflare discussed by \cite{Zanna_2022ApJ} only reached 6 MK. That microflares can have peak temperatures as small as 4 MK has already been
pointed out with a multi-wavelength study by \cite{mitra-kraev_delzanna:2019}.

Various classes of microflares have been studied using different instruments such as 
RHESSI (\citealt{2008ApJ...677.1385C} and \citealt{2008ApJ...677..704H}),
STIX (\cite{2021A&A...656A...4B}),
NuSTAR (\citealt{2021ApJ...908...29D} and \citealt{2021MNRAS.507.3936C}), and
MinXSS (\citealt{2017ApJ...835..122W}; \citealt{2018SoPh..293...21M}). All these papers have reported a range of microflare temperatures,  some of which exceeded 10 MK. The \citealt{2017ApJ...835..122W} and \citealt{2018SoPh..293...21M} studies related to C-class and M-class flares, which are a little larger than the B-class microflare we study in this paper. These utilized MinXSS observations, an instrument similar to XSM. Combined observations of small flares with modelling were studied by \cite{2018ApJ...856..178P}.
In this paper, we carry out a comprehensive investigation of the peak temperature of a B1.8$-$ flare, using EUV and X-ray observations by SDO/AIA, XRT/Hinode, and XSM/Chandrayaan-2. Using a time-resolved spectroscopic analysis of XSM spectra, we also provide some insights into the changes in chemical abundances during this microflare. 
Compared to our previous study (\citealt{Zanna_2022ApJ}), not only do we use a different 
XRT filter ratio, but the flare clearly has hotter emission as 
the flare loops were visible in the AIA 94 and 131~\AA\ band and the high temperature iron feature around 6.6 KeV was clearly visible in the XSM spectra during the peak flare phase.

 
 Section \ref{sec:obs} provides an overview of the observations, with details of the AIA observations in subsection \ref{subsec:aia}; the XSM observations in \ref{subsec:xsm} and the XRT observations in \ref{subsec:xrt}. The results of the combined observations from all the three instruments are discussed and compared in Section \ref{sec:multi}. The findings are summarized in Section \ref{sec:con} with further discussion and the conclusions of the paper.

\begin{center}
\begin{figure*}
\hspace{-2.0cm}
\mbox{
\includegraphics[width=0.6\linewidth,angle=180]{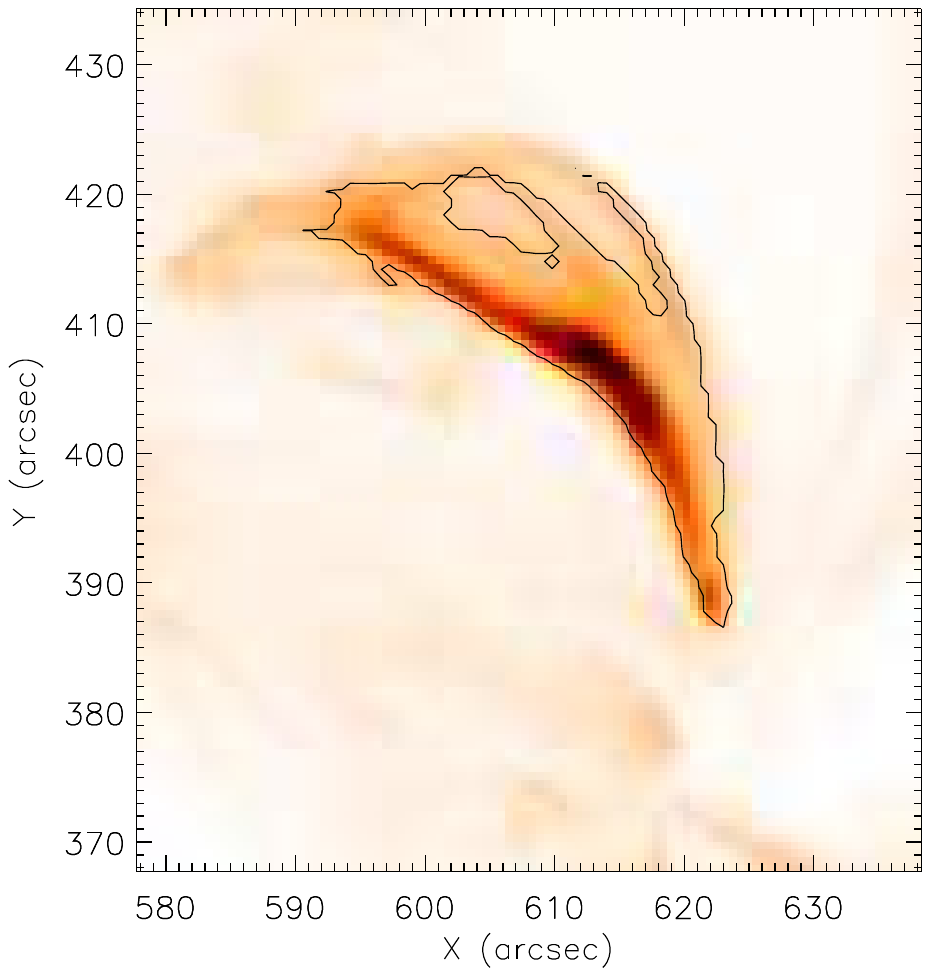}
\hspace{-2.0cm}
\includegraphics[width=0.6\linewidth,angle=180]{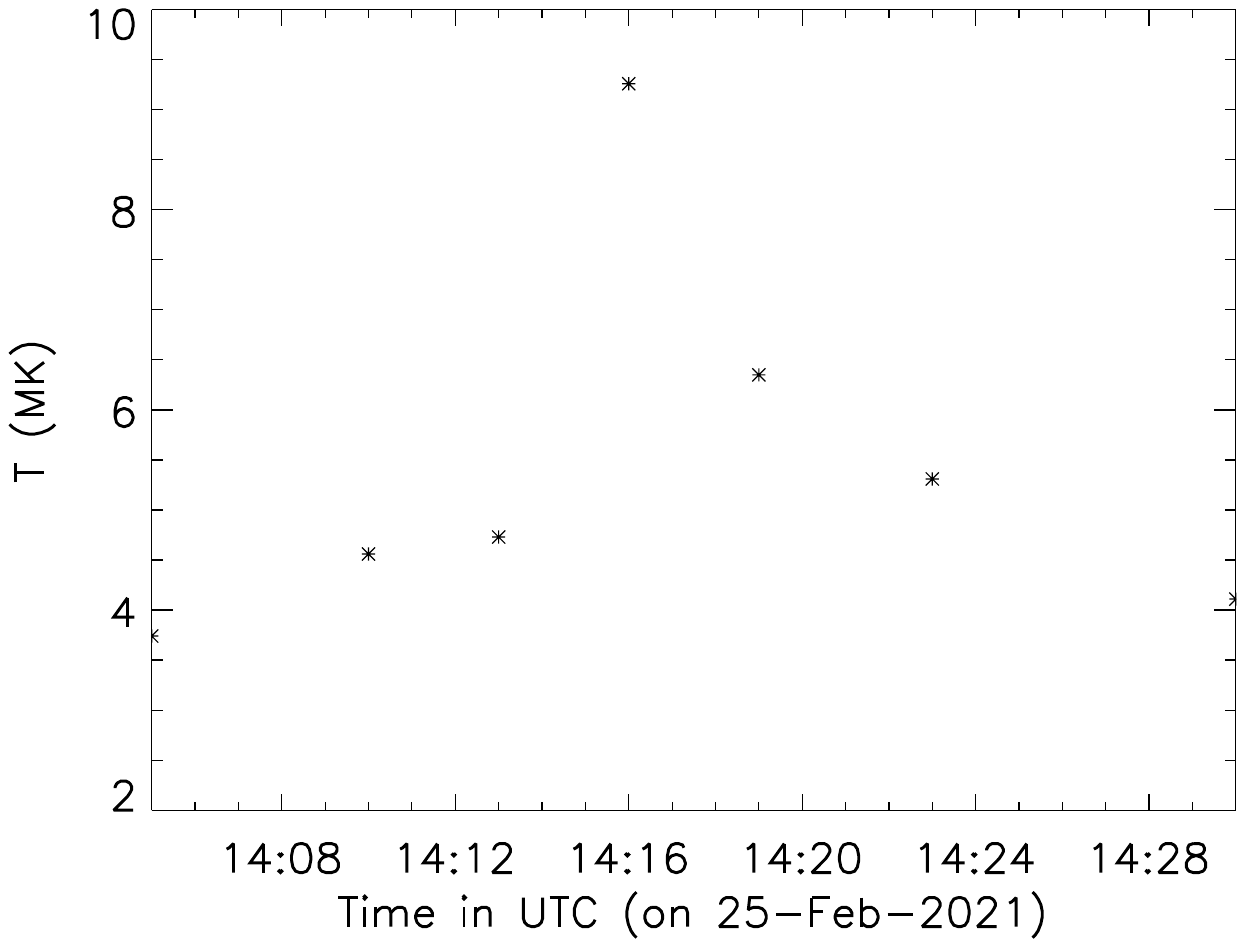}
}
\caption{Left panel: The AR in the 131\AA wavelength of SDO/AIA, showing the bright loop structures. The black contour is overlaid for the emission measure at log T =7.1 showing temperatures above 9MK. Right panel: The temporal evolution of the average effective temperatures integrated over 3 mins intervals from the contour shown in the left panel.}
\label{fig:contour}
\end{figure*}
\end{center}

\begin{center}
\begin{figure*}
\includegraphics[scale=0.7]{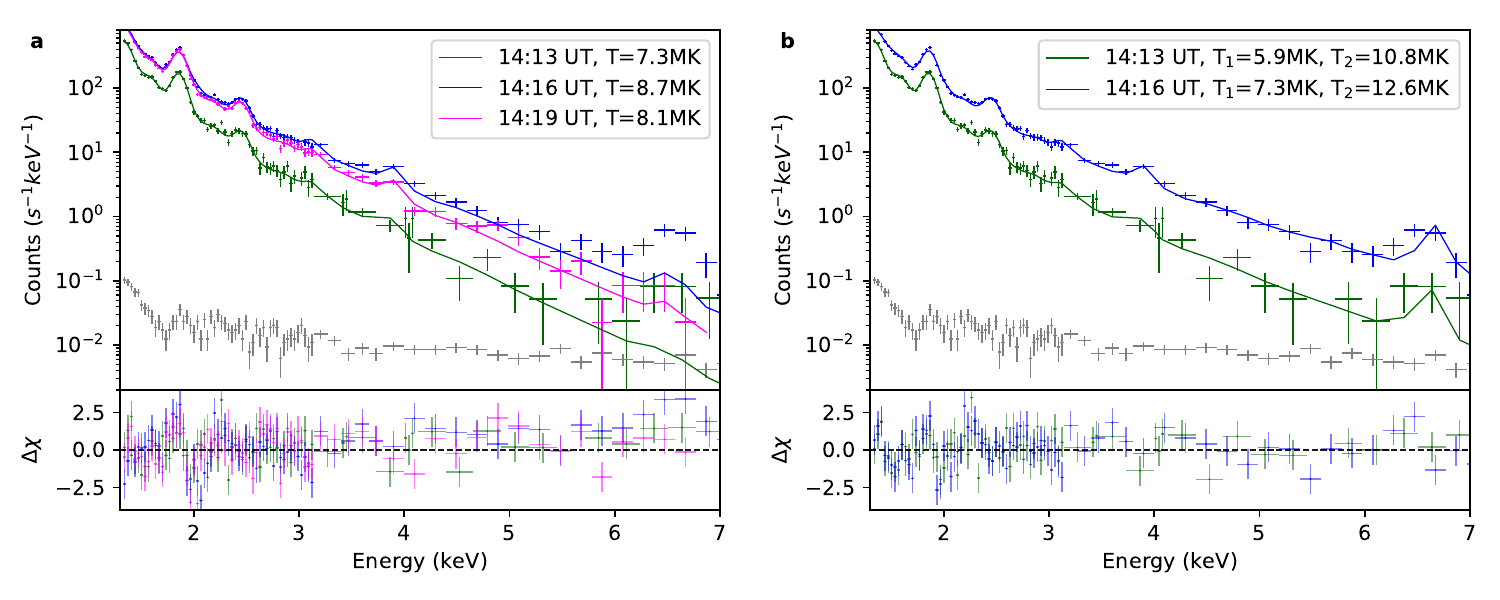}
\caption{XSM observed spectra (points with error bars) modeled with two-temperature (panel a) and three-temperature (panel b) models (solid lines). Different colors represent the spectra at different times during the flare, as mentioned in the plot. The temperature labels in the plot indicate the best-fitted temperatures associated with the flaring plasma (see the text for details).}
\label{fig:XSM_fittedSpec}
\end{figure*}
\end{center}

\section{Observations and Results} \label{sec:obs}

\subsection{AIA} \label{subsec:aia}
The Atmospheric Imaging Assembly (AIA) is an imaging instrument onboard the Solar Dynamics Observatory (SDO). Its multi-wavelength channels cover different regions of the atmosphere of the Sun from the photosphere to the corona, with a large temperature range from log T = 3.7 to 7.3. Images are recorded every 12s for the EUV  channels (94, 131, 171, 193, 211, 335 \AA).

The full field-of-view covers 4096$\times$4096 pixels in both x and y directions where every pixel has a resolution of 0.6$\arcsec$. In this study, we analyze a B-class flare that occurred on February 25th, 2021. The Active Region (AR) 12804, which includes both coronal loops and the flaring region, is the focus of our study. The region of interest spans from 440$\arcsec$ to 740 $\arcsec$ in the x-direction and from 260$\arcsec$ to 560$\arcsec$ in the y-direction and is shown in Figure \ref{fig:aia}. The corresponding filter wavelengths for each panel are noted at the top. 
The flare lasts from 14:00 UT to 14:35 UT, with its peak emission occurring at 14:16 UT. 
The hotter channels, such as 94~\AA\ and 131~\AA, show bright hot loops within the core of the AR during the peak of the flare, as well as cooler and longer coronal loops which are also clearly visible in the other channels.

\begin{figure}
\centering
\includegraphics[scale=0.7,width=7.0cm,height=12cm]{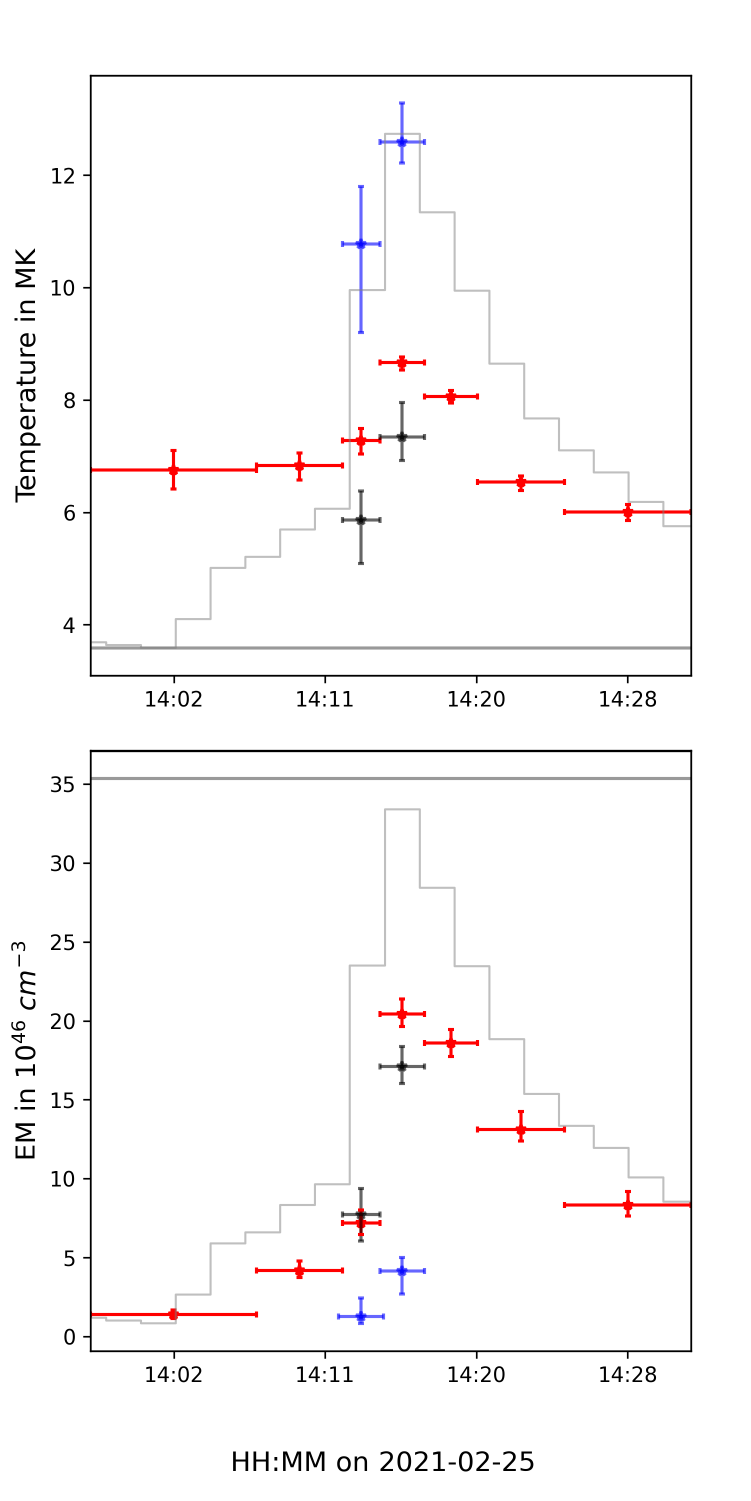}
\caption{The evolution of the temperature (upper) and EM (lower) during the flare as obtained from the spectral analysis of XSM data. The red points represent the parameters associated with the flaring plasma for a 2T analysis (see text). Whereas the grey horizontal line represents the parameters associated with the non-flaring background plasma. The blue and black points near the peak of the flare represent the parameters obtained from the 3T model. The grey curve represents the flare light curve.}
\label{fig:XSM_T_EM}
\end{figure}

\begin{figure*}
\centering
\includegraphics[scale=0.4,angle=0,width=18cm,height=20cm,keepaspectratio]{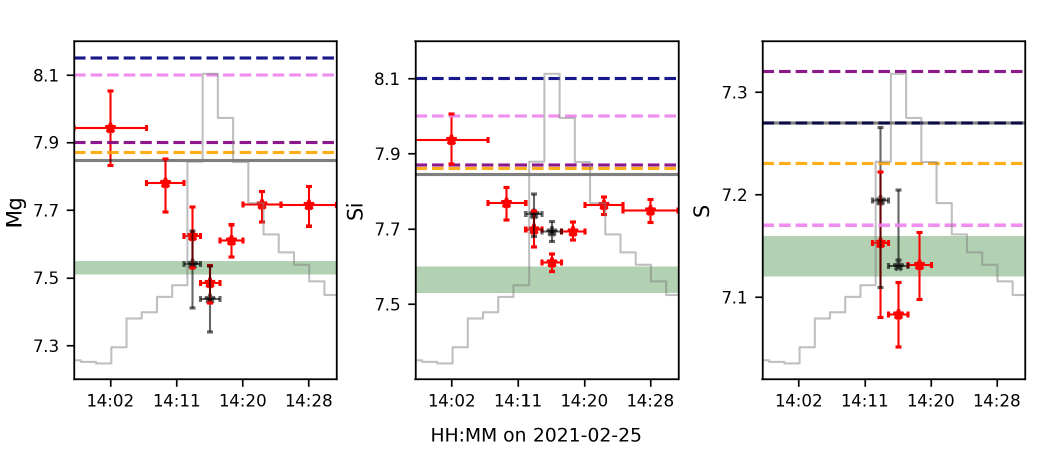}
\caption{Evolution of the abundances (points with error bars) of Mg, Si, and S during the flare. The red and black points represent the abundances obtained from 2T and 3T analyses (see text for details). The horizontal black line represents the abundance of the non-flaring plasma. The horizontal dashed lines of different colors represent the reported coronal abundances by \cite{1992PhyS...46..202F} (blue), \cite{Fludra_1999} (purple), and \cite{Schmelz_2012} (orange) as compiled in the CHIANTI database. The magenta line represents the values reported by \cite{delzanna:2013}. The green shaded region shows the range of reported photospheric abundance in the literature. The grey curve represents the flare light curve.}
\label{fig:XSM_Abundance}
\end{figure*}

We performed a differential emission measure (DEM) analysis using a sparse inversion code written by Mark Cheung \citep{2015ApJ...807..143C} on the full-disc data of SDO/AIA at the flare peak time, 14:16 UT. Figure \ref{fig:dem} displays the emission measure (EM) at different temperatures for the zoomed-in region indicated by  the black box in the top left panel of Figure~\ref{fig:aia}.  %
 
We set a minimum temperature of log T [K]= 5.8, with 16 bins of size 0.1 as an input. The maximum temperature for the DEM reconstruction reached log T =7.3. We have used the maximum temperature to be log T [K] = 7.3 for our work because we do not have temperature constraints at higher temperatures. The EM maps show typical emission from the active region around log T = 6.5 (represented by red and green colors). The map at the bottom right corner, covering the temperature range of log T = 6.9 to 7.1, shows high emission measure (in red) for the small flare studied in this analysis.
The black points indicate data dropouts where the EM inversion failed.  These bad data points do not affect any of our results. The same zoomed-in view of the boxed region of AR in the 131~\AA\ channel of SDO/AIA and contour of emission measure at log T = 7.1 (approximately 12 MK) is displayed in the left panel of Fig. \ref{fig:contour}. 
The right panel shows the temporal evolution of the effective temperatures (averaged values weighted by the emission measure) for 3-min averaged data from locations indicated by the black contour in the left panel. The effective temperature reaches a peak of 9.8 MK at 14:16 UT.

\begin{center}
\begin{figure*}
\hspace{-1.0cm}
\includegraphics[scale=0.4,angle=0,width=10cm,height=10cm,keepaspectratio]{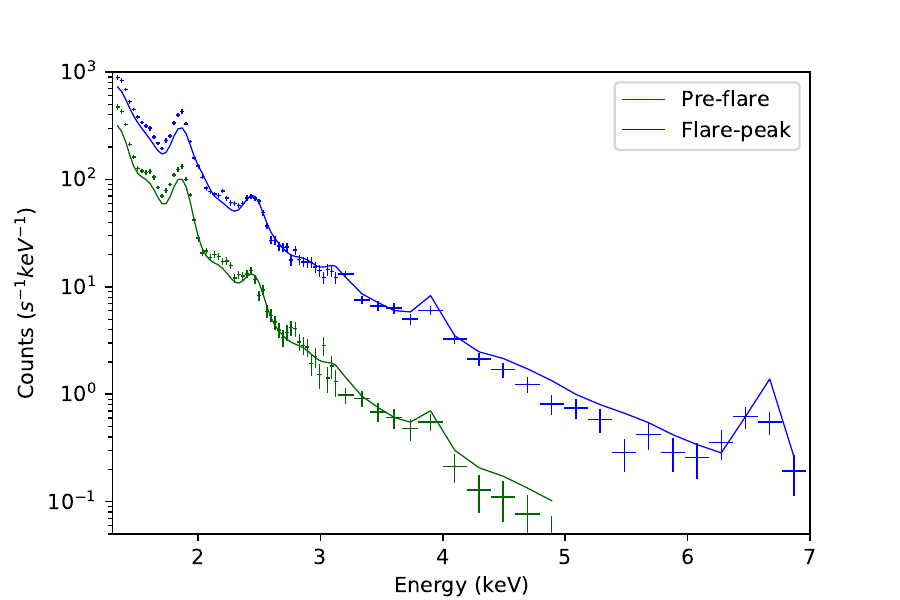}
\hspace{-1.0cm}
\includegraphics[scale=0.6,angle=0,width=9cm,height=9cm,keepaspectratio]{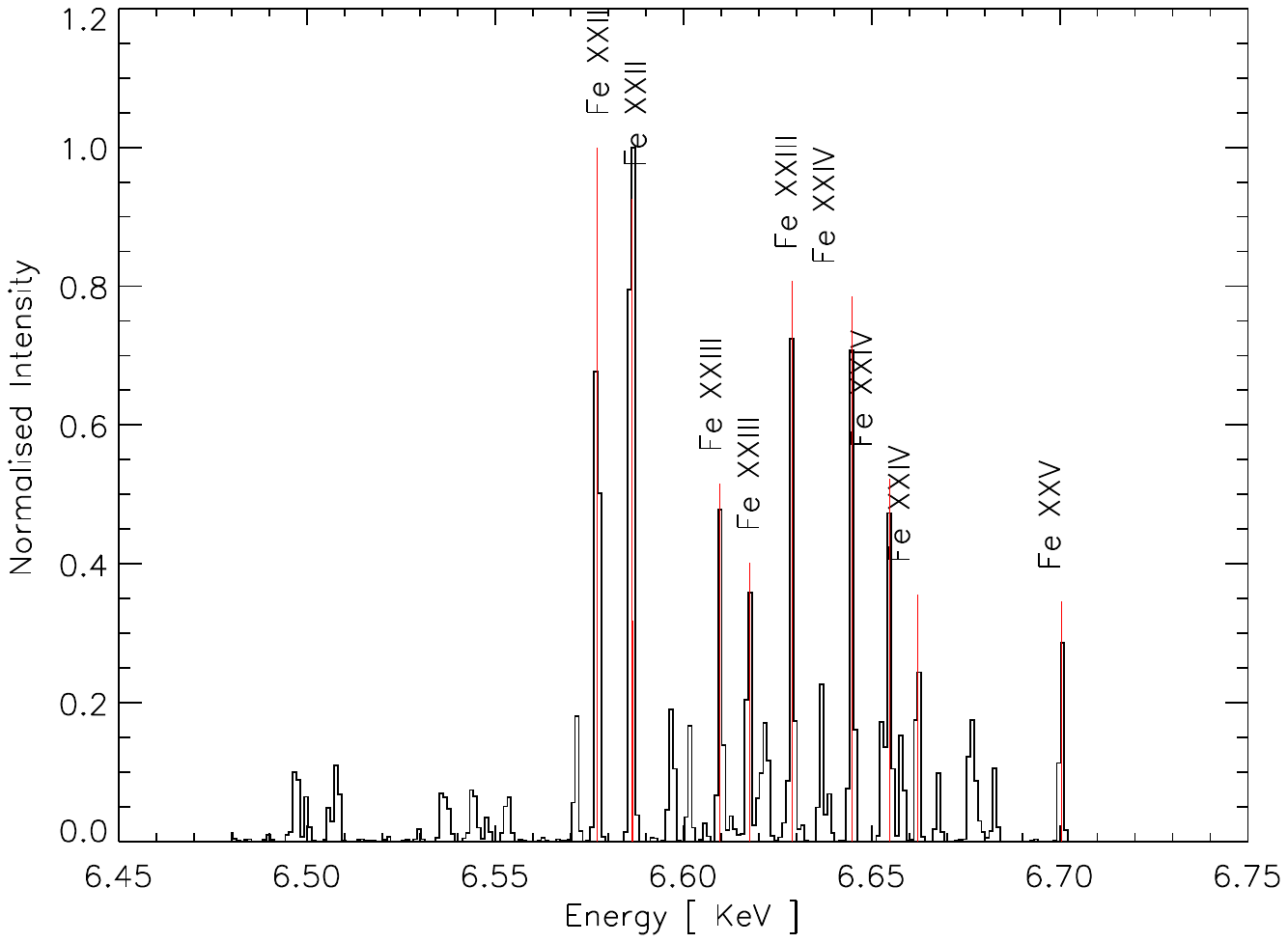}

\caption{Left panel: XSM observed spectra (points with error bars) overplotted with the predicted spectra (solid lines) obtained from the AIA EM for pre-flare (14:13 UT) and flare peak (14:16UT) time. Since the temperatures and EMs derived from XSM and AIA are slightly different, due to differences in their energy sensitivities, a small deviation between the observed and predicted spectra at higher energies is reasonable. Right panel: Simulated spectra for the AR on 25th February 2021 using EM values from XSM at flare peak time.}
\label{fig:simulatedXSMspec}
\end{figure*}
\end{center}

\subsection{XSM} \label{subsec:xsm}
XSM observed this small flare in a disk-integrated mode when no other flaring activity was present on the solar disk. The XSM data is available from ISRO Space Science Data Archive (ISDA\footnote{https://pradan.issdc.gov.in/ch2/}). 
We used the level-1 XSM data and processed them  to level-2 using the XSM Data Analysis Software (XSMDAS:~\cite{xsm_data_processing_2020}).
As we are interested in the time-resolved spectral analysis, we have generated the XSM spectrum at multiple time bins during the flare, following the standard procedure as discussed in ~\cite{Mondal_2021ApJ}.

For the XSM spectral analysis, we use the \verb|chisoth| model~\citep{Mondal_2021ApJ}, which is a local model in X-ray Spectral Fitting Package (XSPEC:~\cite{ref-xspec}). 
The \verb|chisoth| uses the CHIANTI atomic database to calculate the theoretical spectrum for a given temperature, volume emission measure (EM), and with the abundances of the elements from Z=2 to Z=30. 
As XSM is observing the Sun in a disk-integrated mode, the observed spectrum during the flare contains the emission from both the flaring plasma along with the non-flaring plasma from the solar disk.
To eliminate the non-flaring part of the overall spectrum, a possible approach is to subtract the non-flaring spectrum obtained before the flare. However, since the XSM's effective area varies over time, it is not advisable to subtract the pre-flare spectrum for spectral analysis. Therefore, we modeled the pre-flare spectrum with an isothermal model to pre-calculate the plasma parameters.
After that, to model the flaring spectra, we have considered two-temperature (2T) components. The first component represents the emission from the flaring plasma, whose temperature, EM, and the abundances of Mg, Si, and S (these line complexes are prominent in the observed spectra) are treated as free parameters.
The second component represents the emission from the non-flaring plasma from the rest of the solar disk, whose parameters are kept fixed during the spectral analysis. 
Figure~\ref{fig:XSM_fittedSpec}a shows the modelled spectra for three-time bins. We found that during the peak phase of the flare, the modelled spectrum (blue curve) was not able to explain the higher energy Fe line complex (at 6-7 keV). 
Similar results were found for the bigger C class flare as reported by \cite{Mithun_2022} and 6.5 keV feature is not being fitted properly.
This indicates that a single temperature component for the flaring plasma in the 2T model is insufficient to explain the observed spectrum at lower and higher energies. Thus in the next step, we have added a third isothermal component in the 2T model and created a 3T model, where one component represents the non-flaring plasma emission, like the 2T model, and the other two temperatures represent the emission from the flaring plasma.  Here we consider the T and EM for both the component as free parameters, whereas the abundances of Mg, Si, and S were considered free parameters after being tied for both the components. The fitted results for the 3T model during the peak of the flare are shown in Figure~\ref{fig:XSM_fittedSpec}b. We found that the observed spectrum is well-fitted with the 3T model.

\begin{center}
\begin{figure*}
\mbox{
\hspace{-2.0cm}
\includegraphics[width=0.5\linewidth,angle=180]{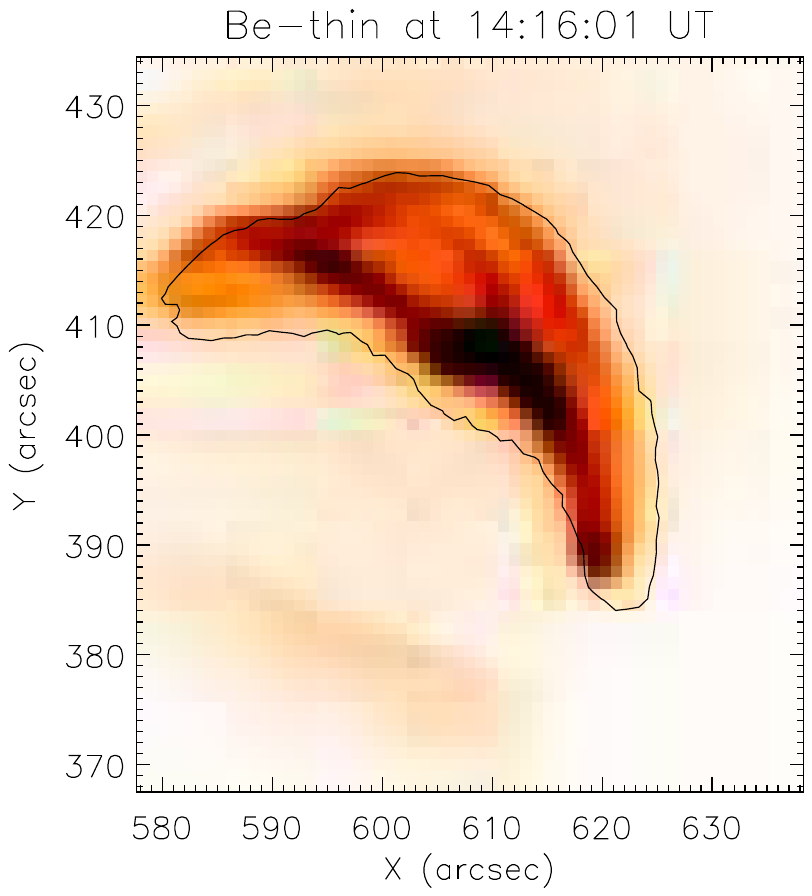}
\hspace{-4.0cm}
\includegraphics[width=0.5\linewidth,angle=180]{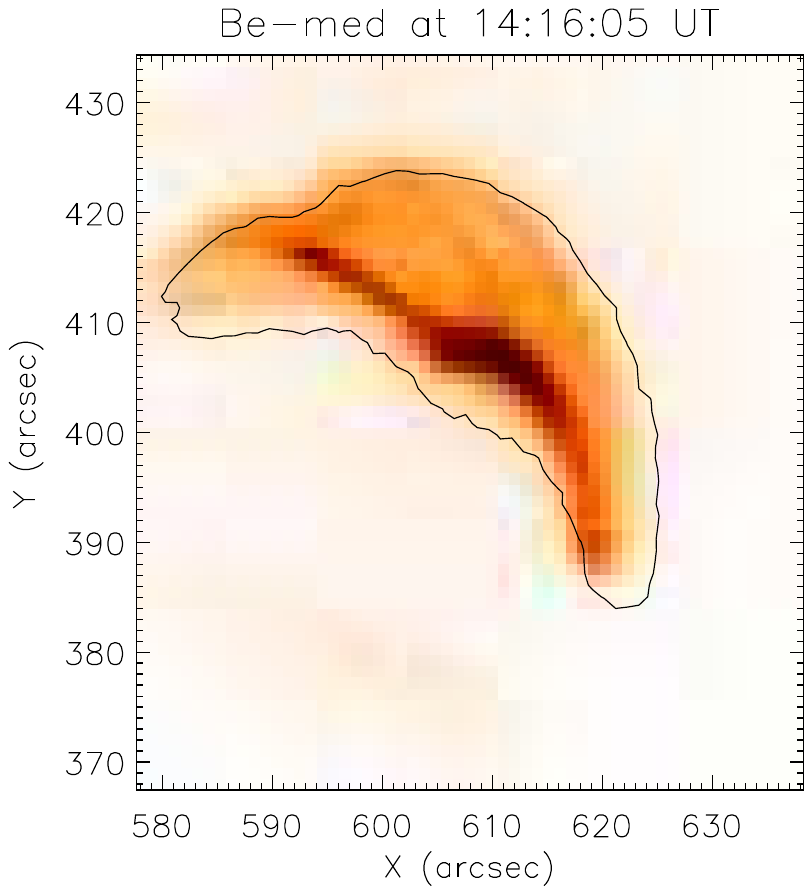}
\hspace{-4.0cm}
\includegraphics[width=0.5\linewidth,angle=180]{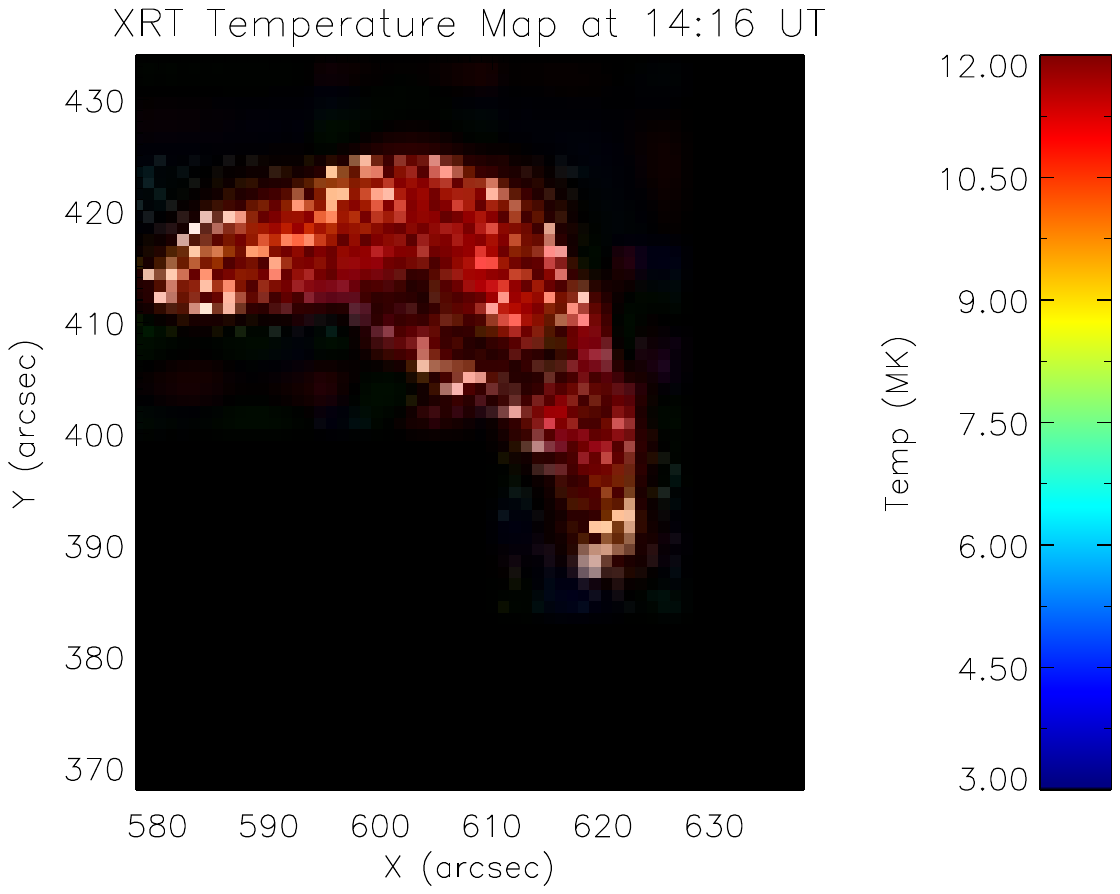}
}
\caption{Left and middle panels: XRT negative images for the  Be-thin and Be-med filters at the flare peak time.  The black contour shows the region considered for the calculation of the averaged effective temperatures. Right panel: XRT effective temperature map at the flare peak time using coronal abundances.}
\label{xrt:map}
\end{figure*}
\end{center}
The results are shown in Figure~\ref{fig:XSM_T_EM} and \ref{fig:XSM_Abundance}. The red points are the best-fit parameters for the flaring component, and the horizontal lines in grey represent the fixed pre-flare parameters for the 2T model. The blue and black points in Figure~\ref{fig:XSM_T_EM} show the best-fitted parameters during the peak of the flare by considering two temperatures for the flaring plasma in the 3T model. Fig. \ref{fig:XSM_Abundance} shows the abundances for different elements Mg, Si,  and S in the left, middle, and right panels respectively. The red and black points indicate abundance values obtained by 2T and 3T model. The variation of abundance during the evolution of the flare for all the elements shows values close to photopsheric range (green shade region) at the flare peak time. 
Note that these abundances are measured from line-to-continuum measurement, whose absolute values are slightly dependent on the choice of the abundances of other abundant elements (whose emission lines are not visible in the XSM spectrum); primarily oxygen, as discussed in Figure~12 of \cite{Mondal_2021ApJ}.
In our present study, we have chosen \cite{1992PhyS...46..202F} abundances for oxygen and other elements.

In order to check the agreement  between the XSM and AIA observations, we have forward-modeled the XSM spectrum using the EM obtained from AIA and compared this with the observed spectrum. 
During the pre-flare and flare-peak, we derived the EM distribution of the full Sun using the images observed by different AIA filters. 
Selecting the bright regions at 3 MK, we have estimated the volume EM of the full disk. Using this volume EM along with the XSM measured abundances, we have modeled the observed XSM spectrum.
The left panel of Figure~\ref{fig:simulatedXSMspec} shows the observed (data points) and the model (solid lines) spectra for the pre-flare (green) and flare peak (blue) duration.
Considering the different temperature sensitivity of XSM and AIA, the agreement is remarkable. 

One interesting question is what contributes to the third hotter 
component in the XSM peak flare spectrum. Usually, the line complex around 6.7 KeV as shown in the left panel of Fig. \ref{fig:simulatedXSMspec}
is assumed to be associated with emission from Fe XXV plus satellite lines, which are 
formed at temperatures in the range 12-20 MK. However, in the case of the microflare we have studied, the temperatures didn't reach as high as 12-20 MK, but we did get a contribution from other lower temperature ions like Fe XXII, FeXXIII, and FeXXIV. This feature has a peak emission from these ions at around 6.6 KeV, indicated by the CHIANTI simulated spectrum using EM values from XSM as shown in the right panel of Fig. \ref{fig:simulatedXSMspec}. It shows that the peak emission comes from 
Fe XXII (around 6.6 KeV), with very little emission from Fe XXV. We can therefore be confident that this hot emission is produced by the flare loops as seen in the AIA 131~\AA\ band, dominated by Fe XXI emission.


\subsection{XRT} \label{subsec:xrt}
The X-Ray Telescope (XRT) on Hinode operates in the 6-60 \AA~ wavelength range and is sensitive to temperatures around log T = 7.0. It consists of multiple channels including Al-mesh, Al-poly, C-poly, Ti-poly, Be-thin, Be-med, Al-med, Al-thick, and Be-thick, which primarily detect emission from the solar corona. 

\begin{center}
\begin{figure*}
\hspace{1.0cm}
\includegraphics[width=0.8\linewidth,angle=180]{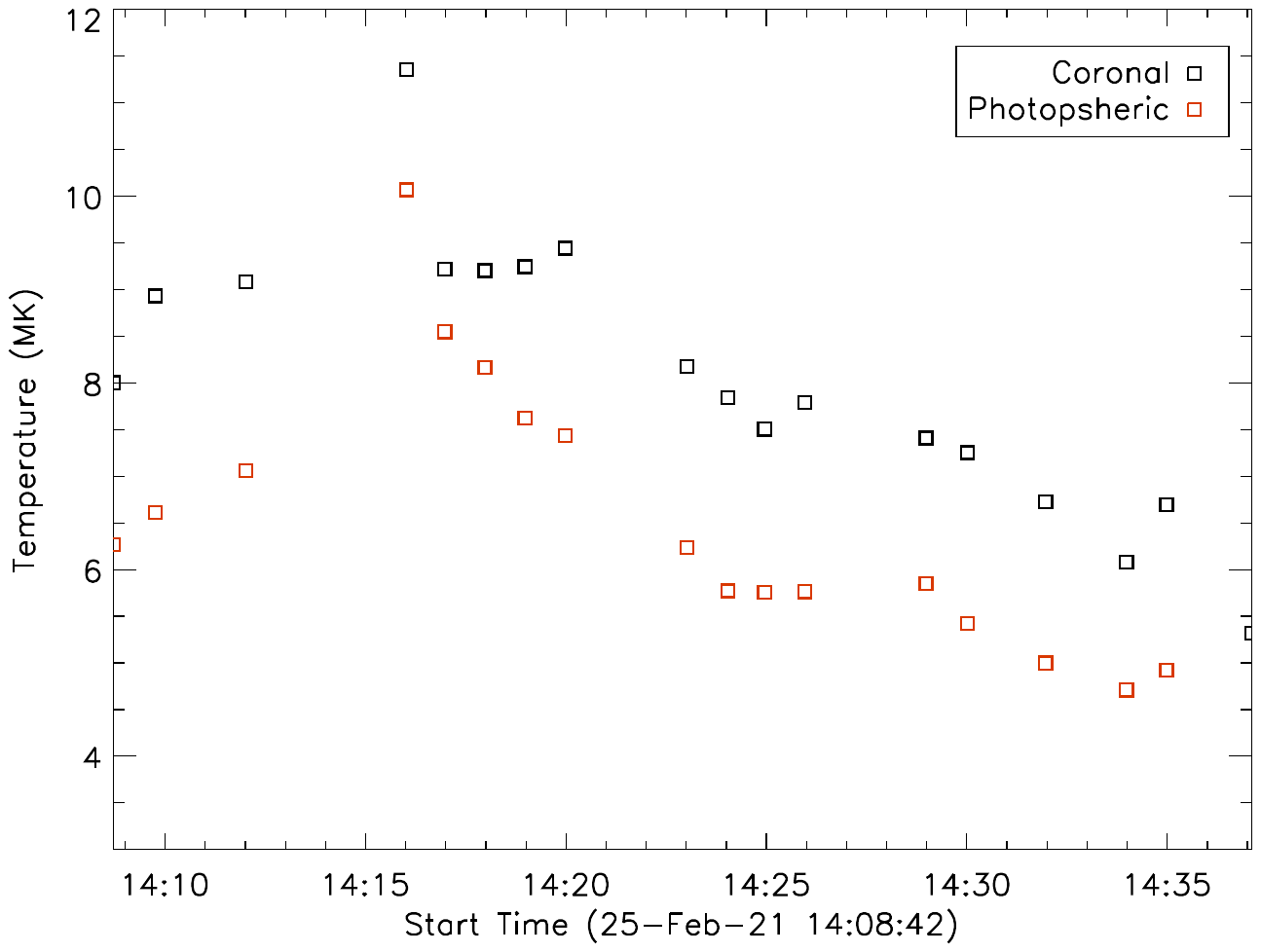}
\caption{The effective temperature variation using Be-thin/Be-med filter ratios from XRT. The black and red color indicates the temperature values using coronal and photospheric abundances, respectively.}
\label{xrt:temp}
\end{figure*}
\end{center}

For our analysis, the XRT files were processed to remove saturation and were normalized using xrt\_prep.pro. The left and middle panels of Fig. \ref{xrt:map} display the XRT images for the Be-thin and Be-med filters at the flare peak time. The FOV is the same as the black boxed region shown in Fig. 1. We then calculated the temperatures weighted by the EM values, within the isothermal assumption. The corresponding effective temperature map in the right panel of Fig. \ref{xrt:map} shows that the core loop structures of the active region are dominated by high temperatures around 10 MK. Fig. \ref{xrt:temp} shows the temporal evolution of the averaged effective temperature, calculated using both photospheric and coronal abundances over the region included within the black contours in the filter images. The black points are the locations outside the intensity threshold range where we couldn’t get meaningful temperatures for temperature calculations and really high temperatures where it shows saturation.

There is a difference of approximately 20 \% between the temperatures obtained using the two sets of abundances. This is due to the different elements contributing to the two XRT filters. Assuming coronal abundances, Be-med filter is dominated by low-FIP elements like Si, Mg, and Fe (Fig. \ref{fig:be_med_phot}). Most of the contributions for Be-thin filter using coronal abundances is also from low-FIP elements like Si, Mg, and Fe  (Fig. \ref{fig:be_thin_phot}). The contribution from high-FIP elements like Ne becomes dominant for photospheric abundances in both the bands resulting in the difference of response ratio using different abundances (Fig. \ref{fig:xrt_temp_ratio}).

In our previous work, we used Be-thin and Al-poly filter combination, but this did not show good agreement in temperatures derived from AIA and XRT, since Al-poly is biased towards lower temperatures. However, the Be-thin filter and Be-med filter is an excellent filter combination to probe high temperatures.

\section{Multi-instrument Results}\label{sec:multi}
\begin{center}
\begin{figure*}
\hspace{3.0cm}
\includegraphics[scale=1]{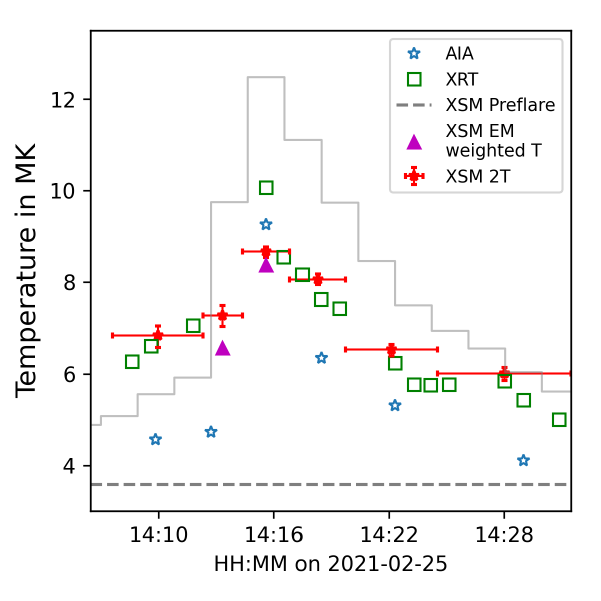}
\caption{The temporal evolution of temperature using AIA (blue asterisk), XRT (green squares), and XSM (red and pink points). The red points represent the flaring plasma temperature obtained from a two-temperature fit of the XSM spectra, where the grey dashed line represents the temperature of the background AR. 
The pink triangular points represent the emission-measure-weighted temperature obtained from the cool and hot components of the flaring plasma while considering three-temperature components to the XSM spectra (see the text for details)
The grey plot indicates the XSM light curve.}
\label{fig:T_xsm_aia_xrt}
\end{figure*}
\end{center}

In this B-class flare, we observed high-temperature plasma, with temperatures reaching around 10 million Kelvin. The combined plot of the temperature evolution from various instruments (Fig. \ref{fig:T_xsm_aia_xrt}) demonstrates good agreement. The grey curve represents the light curve from the XSM, while the blue stars indicate temperatures derived from the SDO/AIA and green boxes show the XRT temperatures calculated using line ratios from the Be-thin and Be-med filters from photopsheric abundance. The black dashed line represents the XSM temperature prior to the flare peak. The red points with error bars indicate the temperature estimated by two-temperature (2T) modeling of XSM spectra. The pink triangular points indicate the emission measure weighted temperatures estimated from 3T modeling of XSM spectra during the peak of the flare, where 2T modeling is insufficient to explain the higher energy ($\sim$6.7 keV) emission in XSM spectra (see Figure~\ref{fig:XSM_fittedSpec},\ref{fig:XSM_T_EM} and the discussion therein).

Before and after the flare peak time, the AIA temperatures are lower than those from XRT and XSM indicating that AIA filters are not sensitive to 5-8 MK temperatures. The Figures  \ref{94images}, \ref{131images}, and \ref{171images} in the Appendix show that the hot flare loops are visible in well structured form in the 94~\AA\ and 131 \AA~ band only at the flare peak time (14:16 UT). At other times before an after the flare, 131 \AA~ shows the features similar to those in the 171~\AA\ channel, indicating cooler emission.
Having only one filter (the 94~\AA) showing hot emission therefore limits the ability to obtain temperatures for the flare loops from AIA outside of the flare peak. Generally, XRT temperature values agree well with XSM as both the X-ray instruments have sensitivity to high temperatures. The AIA temperatures during flare peak also agree well with XSM and XRT. XRT temperatures obtained using photospheric abundances agree more closely with the AIA and XSM results, as expected because we know from XSM analyses that abundances are close to photospheric values during the flare peak.

\section{Conclusions and Discussion}\label{sec:con}

We have observed high-temperature plasma (above 10 MK) at the peak of this microflare (B-class). 
The combined temporal evolution of temperatures from different instruments (XSM, AIA, and XRT) shown in Fig. \ref{fig:T_xsm_aia_xrt} demonstrates excellent agreement, within 20\%, around the peak emission. AIA temperatures are lower than those from other instruments before and after the flare peak emission time. 
This is likely because AIA is more sensitive to the lower temperature plasma than the other two instruments. As shown in the Appendix (reference Fig. \ref{94images} and \ref{131images}), we see the hot loop structure in both 94 \AA~ and 131 \AA~ passbands only around the peak of the flare. At other times, before and after the flare peak, the hot loops are only visible in 94 \AA. The lack of a strong signal in 131 \AA~ produces effective temperatures from the AIA EM results that are significantly lower than those from XRT and XSM.
The agreement between XRT, using the Be-thin to Be-med filters, and XSM is excellent even before and after the peak, better than 10\%. We place most confidence in the agreement of results for temperature for all three instruments during the flare peak. The EM maps using different channels of AIA show emission from temperatures up to 12 MK from a confined region of the loops during the flare peak.

On the basis of the Si abundance variation measured by XSM during the flare, as shown in Fig.~\ref{fig:XSM_Abundance}, we expect the elemental abundances to vary from their photospheric values by less than a factor of 2. Therefore, we do not expect the temperature obtained from XRT to vary by more than $\sim$ 10\%.

 
The combination of Be-thin/Be-med filters in XRT is found to be better for high temperature diagnostics than  Be-thin/Al-poly. We strongly recommend its use in future studies. XSM provides full-disc spectral data, but the XRT capabilities have allowed us to probe the active region and small flare of interest. We found evidence of elemental abundances decreasing towards photospheric values around the flare peak time while remaining at coronal values before and after the peak time, consistent with results previously studied and reported from XSM~\citep{Mondal_2021ApJ,Mithun_2022,laksitha_2022} and MinXSS~\citep{2017ApJ...835..122W}.

These coordinated observations are valuable 
for improving our understanding of the observational capabilities of instruments with different temperature sensitivities. That is how the observations with these different instruments complement each other. However, due to limited coordinated observations, our study is limited to a single event. Compared to our previous multi-wavelength study (\citealt{Zanna_2022ApJ}), this flare of similar class definitely has a much higher peak temperature and is more compact in size.
\cite{2013ApJ...770..126B} observed 17 GOES C-class flares and reported the diversity of events and variation of different parameters within the same class. They also found that shorter duration flares, which are smaller in geometric size and emit less radiation, are the hottest and densest events.
\cite{2021A&A...656A...4B} pointed out that the STIX observations of small class A and B flares indicate that the emission was mainly thermal in nature. The consistency of our peak temperatures from different instruments is significant in this regard. Joint studies of STIX and XSM observations, with combined analysis software, are being developed to better determine the thermal and non-thermal contributions to flares, small and large. 
Coordinated observations of a larger sample of microflares could significantly improve our understanding of these events. These observations and our results are most valuable for comparisons with modelling.

\section{Acknowledgments}

We acknowledge the financial support by STFC (UK) under the Research Grant ref: ST/T000481/1 and the research facilities provided by Department of Applied Mathematics and Theoretical Physics (DAMTP), University of Cambridge.  We acknowledge the use of data from the Solar X-ray Monitor (XSM) on board the Chandrayaan-2 mission of the Indian Space Research Organisation (ISRO), archived at the Indian Space Science Data Centre (ISSDC). XSM was developed by Physical Research Laboratory (PRL) with support from various ISRO centers. KR is supported by contract NNM07AB07C from NASA to SAO.  The collaboration between the PRL and Cambridge groups has been facilitated through the Royal
Society grant No. IES-R2-170199. We thank various facilities and the technical teams from all contributing institutes  and Chandrayaan-2 project, mission operations, and ground segment teams for their support. Research work at PRL is supported by the Department of Space, Govt. of India. We acknowledge the work of the NASA/SDO, AIA, EVE and HMI teams. Hinode is a Japanese mission developed and launched by ISAS/JAXA, with NAOJ as a domestic partner, with NASA and STFC (UK) as international partners. It is operated by these agencies in co-operation with ESA and NSC (Norway). CHIANTI is a collaborative project involving George Mason University, the University of Michigan (USA), University of Cambridge (UK) and NASA Goddard Space Flight Center (USA).  


%


\bibliography{25feb_microflare}{}
\bibliographystyle{aasjournal}

\appendix
\section{Supplementary Information}
\label{appendix:a}
\setcounter{figure}{0}
\renewcommand\thefigure{\Alph{section}.\arabic{figure}}

\begin{center}
\begin{figure*}
\includegraphics[width=1\linewidth,angle=180]{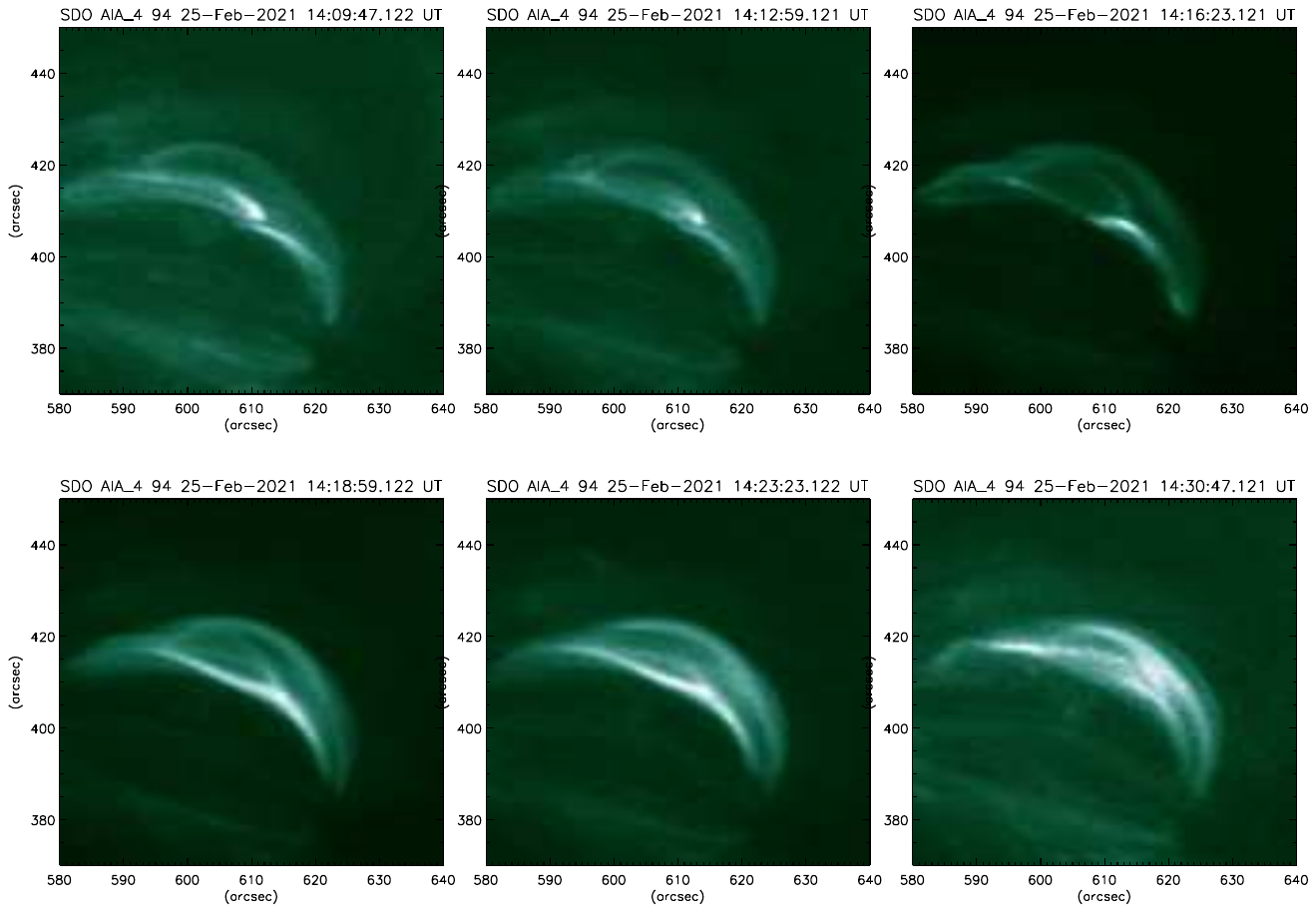}
\caption{Evolution of the loop structures observed in the 94~\AA\ channel of SDO/AIA}
\label{94images}
\end{figure*}
\end{center}

\begin{center}
\begin{figure*}
\includegraphics[width=1\linewidth,angle=180]{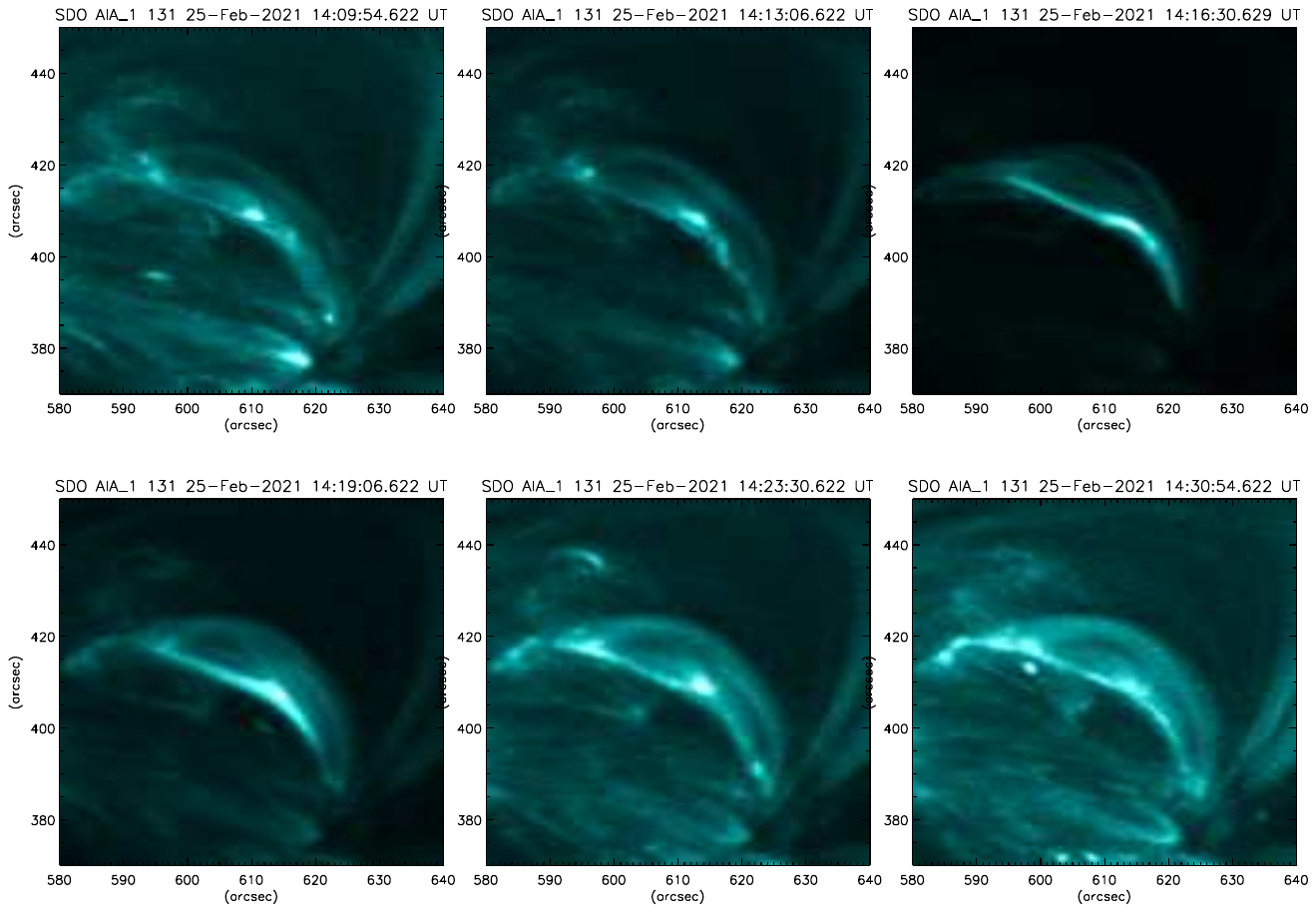}
\caption{Evolution of the loop structures observed in the 131 \AA channel of SDO/AIA}
\label{131images}
\end{figure*}
\end{center}

\begin{center}
\begin{figure*}
\includegraphics[width=1\linewidth,angle=180]{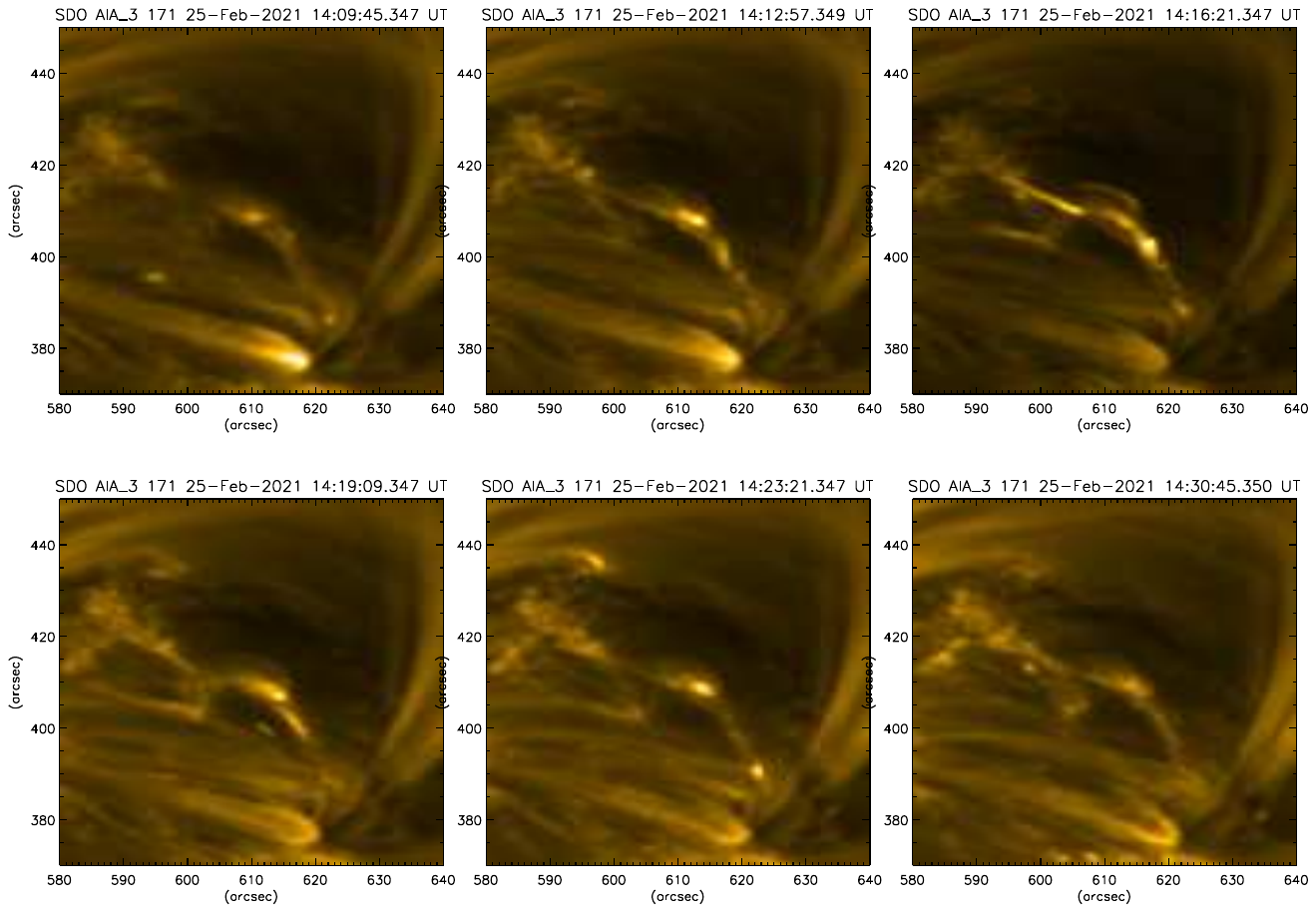}
\caption{Evolution of the loop structures observed in the 171 \AA channel of SDO/AIA}
\label{171images}
\end{figure*}
\end{center}

Figures \ref{94images}, \ref{131images}, and \ref{171images} depict the evolution of the microflare in the SDO/AIA 94, 131, and 171~\AA\ channels. While the AIA temperatures match closely with XSM and XRT at the peak of the flare, they show lower values at all other times before and after the peak. The sequence of different channels reveals the morphological behavior of hot loops at different time intervals. The 94~\AA\ shows the evolution of hot loop structures which are clearly visible at all the times. The hot loop structures are co-spatial and also visible in 131~\AA\ passband during the flare peak time. At other times, the 131~\AA\ passband shows fragmented brigtenings that are co-spatial with the 171~\AA\ channel, which is sensitive to temperatures below 1 MK. This bright emission is mostly coming from the lower temperature regions which is typical of quiet-Sun transition region.

\begin{center}
\begin{figure*}
\includegraphics[scale=0.6,angle=180]{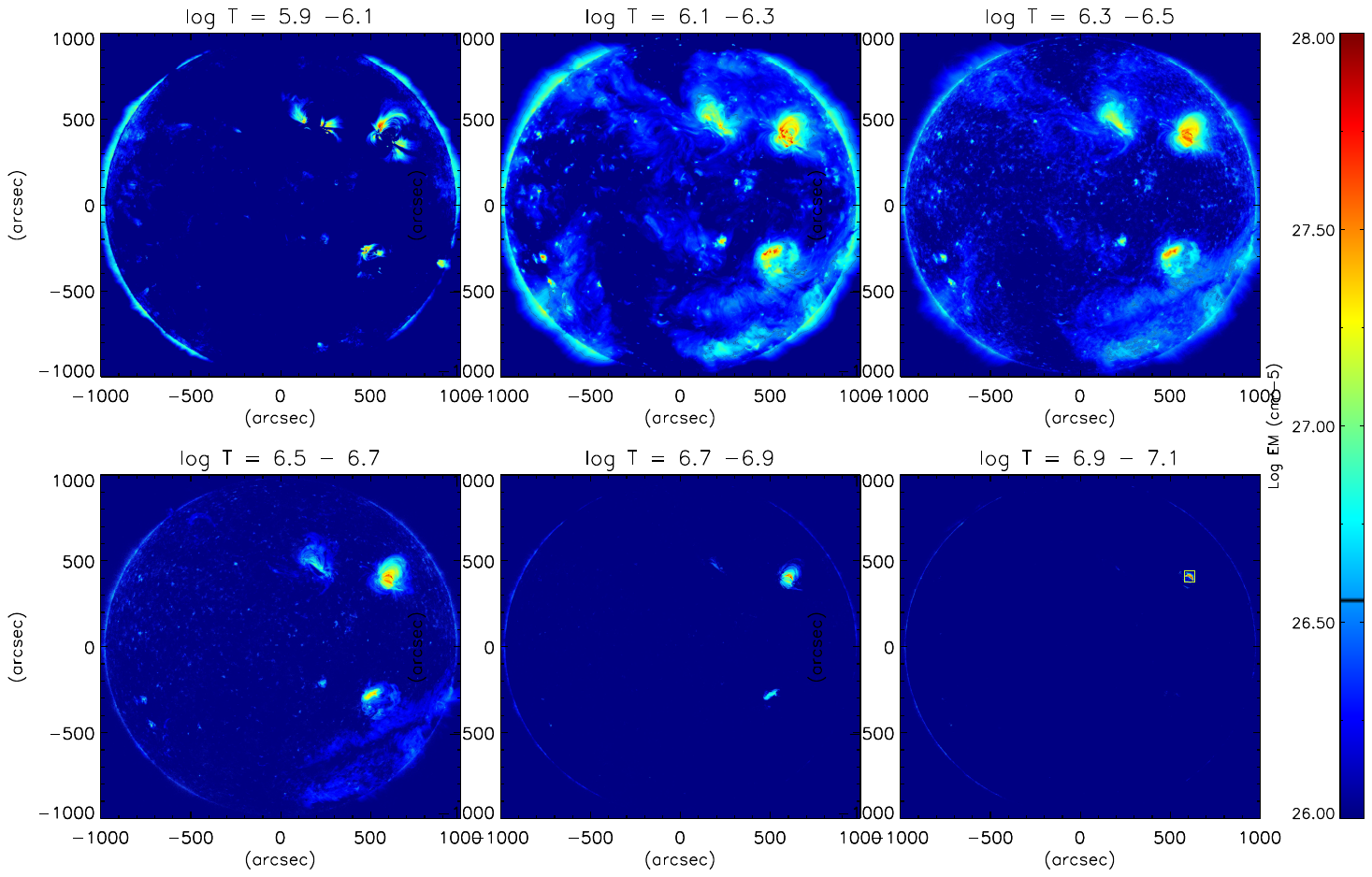}
\caption{Full disc DEM for 3 min averaged data from 14:14 UT to 14:17 UT during the flare peak time.}
\label{fig:dem_fulldisc}
\end{figure*}
\end{center}

\begin{center}
\begin{figure*}
\hspace{-1.0cm}
\mbox{
\includegraphics[width=0.5\linewidth,angle=0]{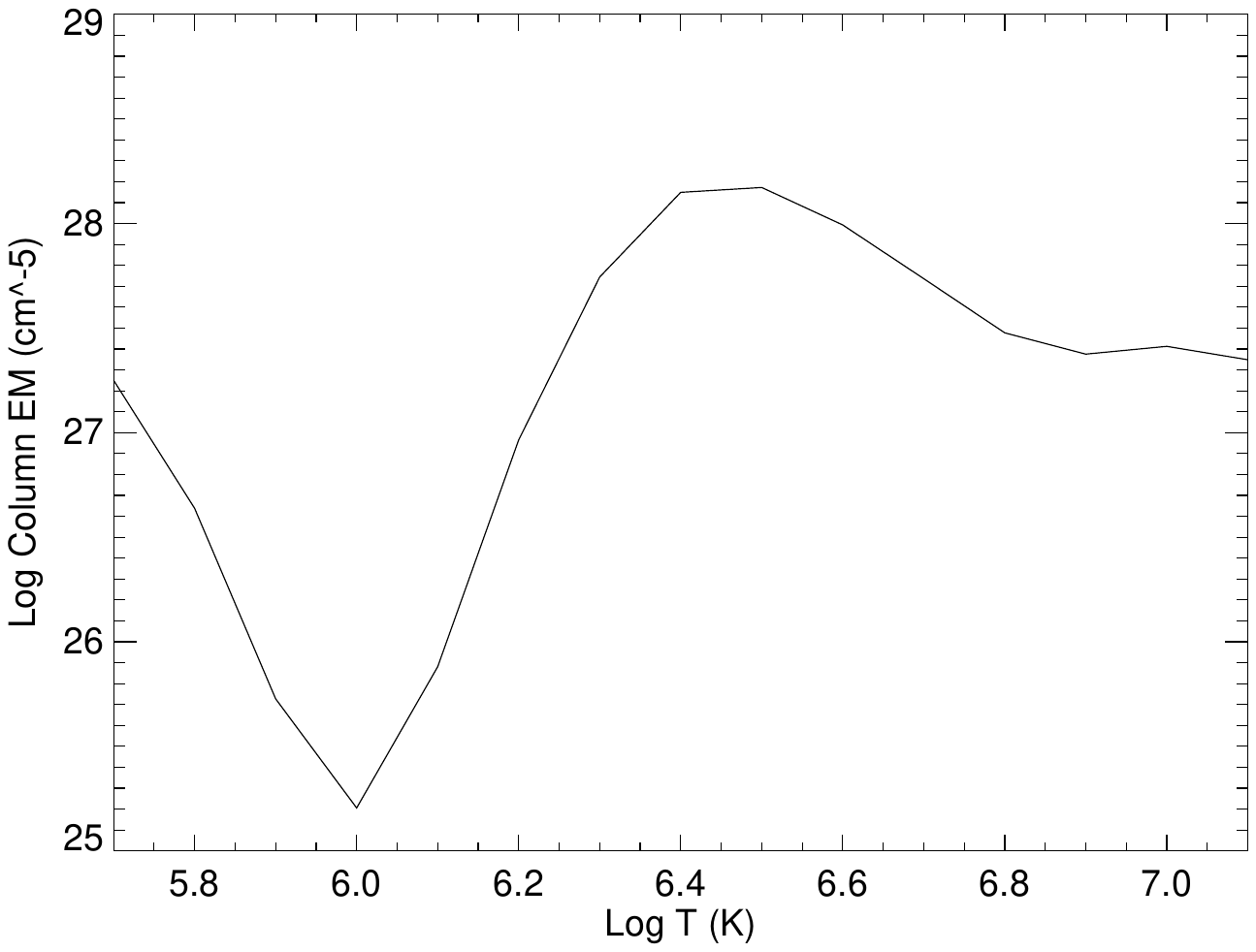}
\includegraphics[width=0.5\linewidth,angle=0]{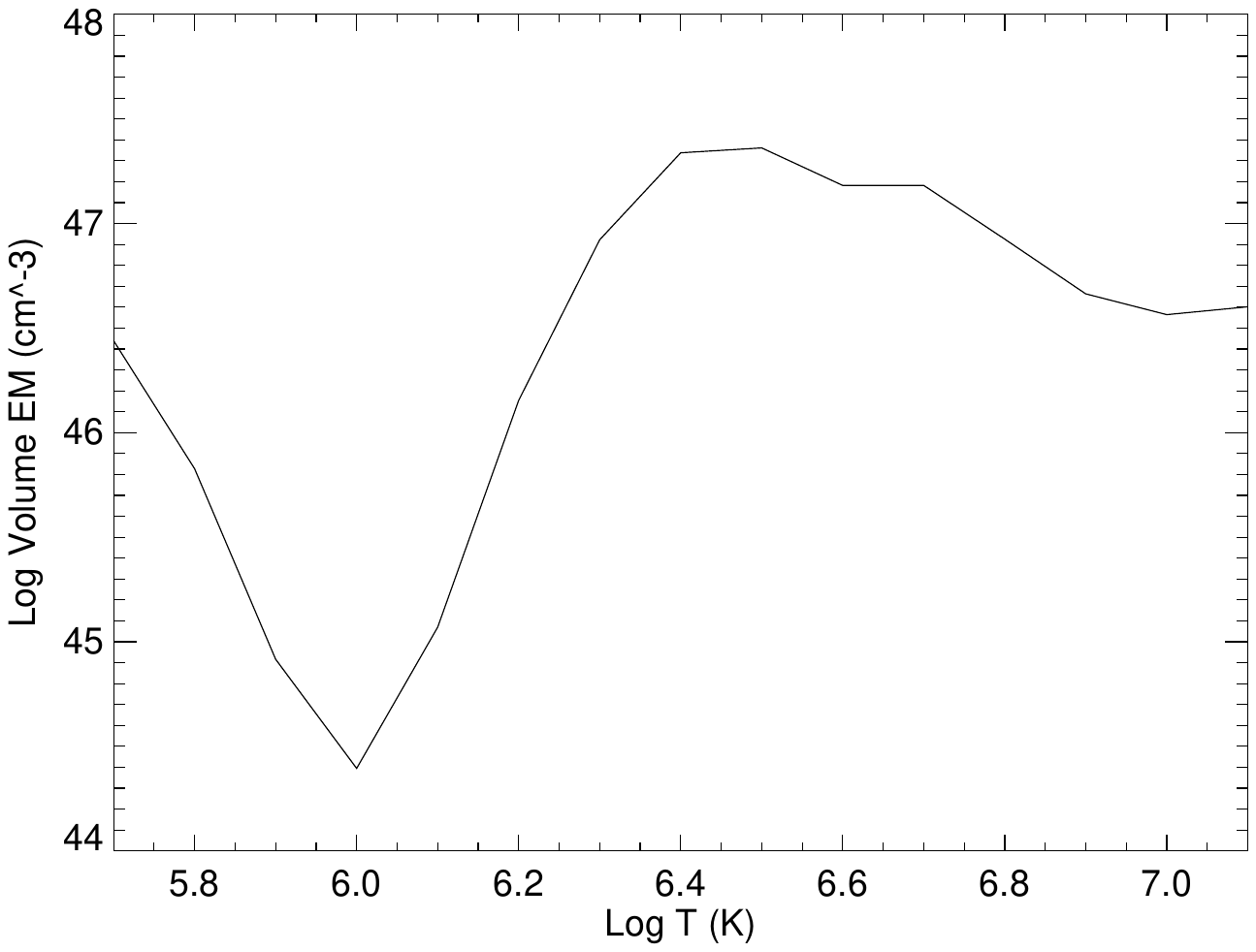}
}
\caption{Left panel: The variation of column EM w.r.t temperature at the flare peak time from the full disc (excluding the limb) averaged over 3min intervals during the flare peak time (14:14 to 14:17 UT). Right panel: The variation of Volume EM w.r.t temperature at the flare peak time from the full disc (excluding the limb) averaged over 3min intervals during the flare peak time (14:14 to 14:17 UT). These values are used as an input to simulate the XSM spectra since the XSM provides full disc emission signal. The size of temperature bins for both the plots is log T [K] = 0.1.}
\label{fig:em_var}
\end{figure*}
\end{center}

Fig. \ref{fig:dem_fulldisc} shows the EM maps for the full disc where the AR of the interest is the one showing emission near the limb at the top right corner. Other small active regions are present as well but don’t contribute significantly at high temperatures above 10 MK. The panels log T/K = 6.5$-$6.7 and log T/K =6.7$-$6.9 show limb brightenings which give overestimated signal in XSM when included. So, we considered the EM for the whole disc, excluding the limb region, to simulate the spectra for XSM. The left and right panels of Fig. \ref{fig:em_var} shows the variation of column and volume EM respectively for the full disc data from SDO/AIA at flare peak time (14:16 UT).

\begin{center}
\begin{figure*}
\hspace{-1.0cm}
\mbox{
\includegraphics[width=0.55\linewidth,angle=180]{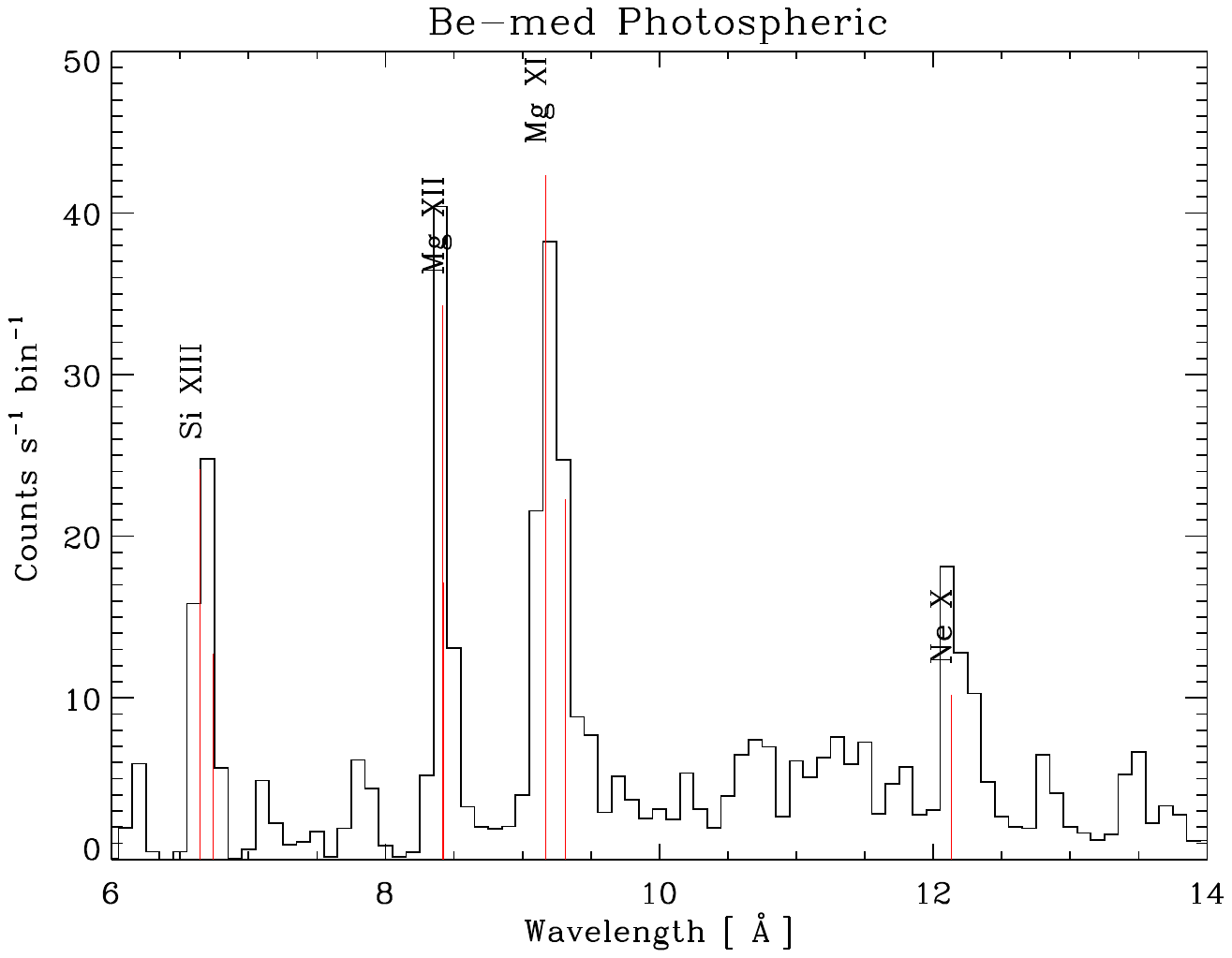}
\includegraphics[width=0.55\linewidth,angle=180]{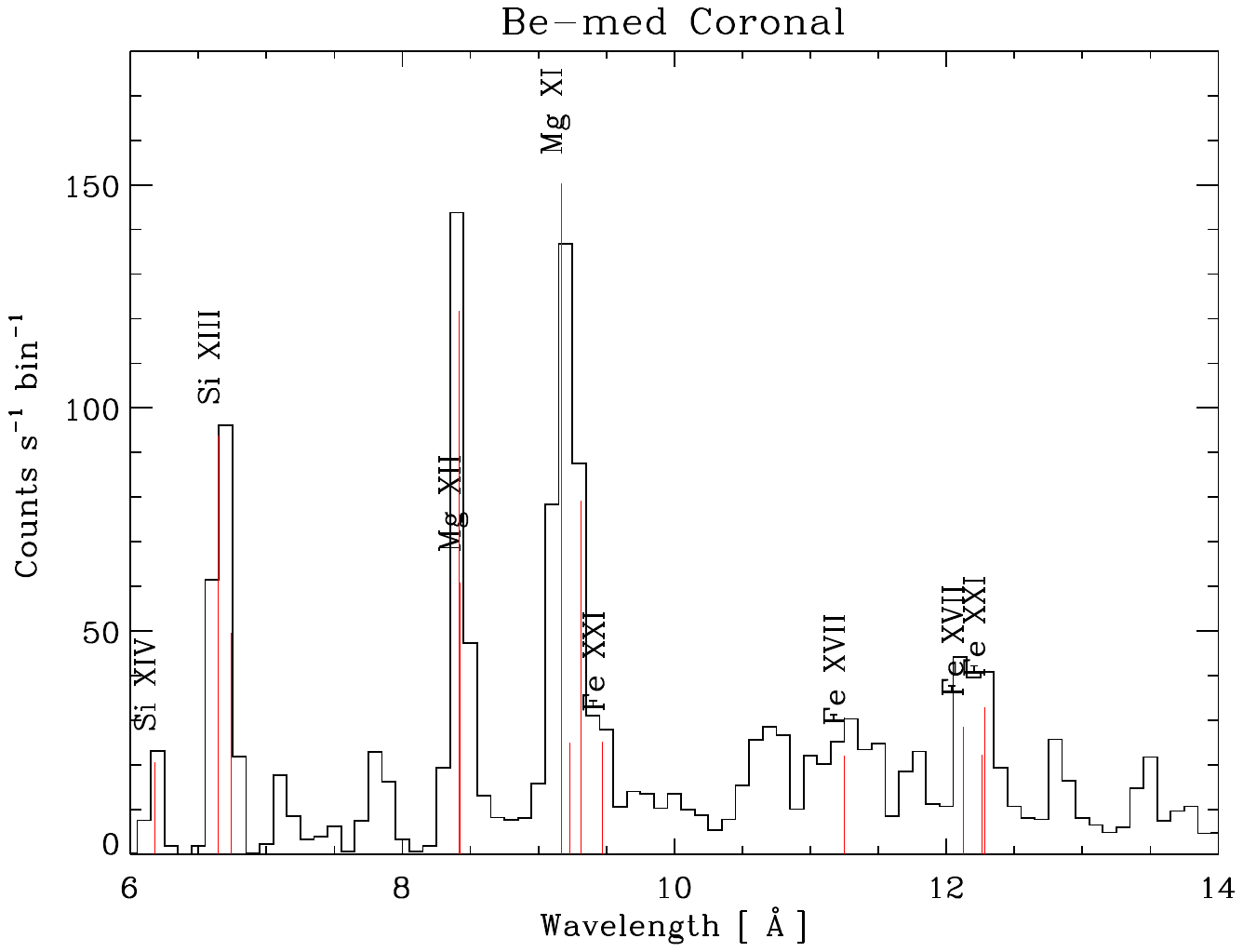}
}
\caption{Left panel: Simulated spectra from CHIANTI corresponding to the Be-med filter of XRT using photospheric abundances. Right panel: Simulated spectra from CHIANTI corresponding to the Be-med filter of XRT using coronal abundances.}
\label{fig:be_med_phot}
\end{figure*}
\end{center}

\begin{center}
\begin{figure*}
\hspace{-1.0cm}
\mbox{
\includegraphics[width=0.55\linewidth,angle=180]{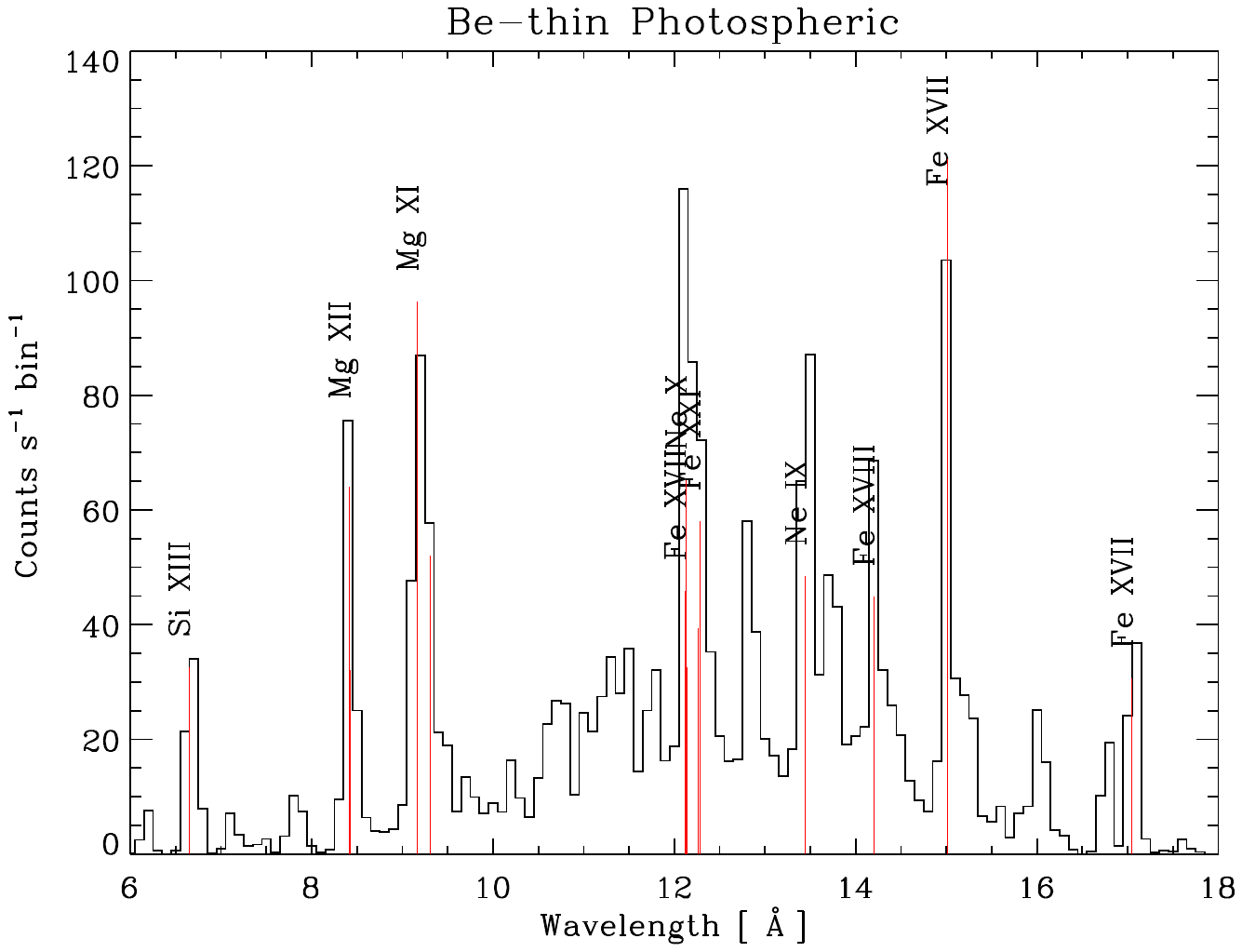}
\includegraphics[width=0.55\linewidth,angle=180]{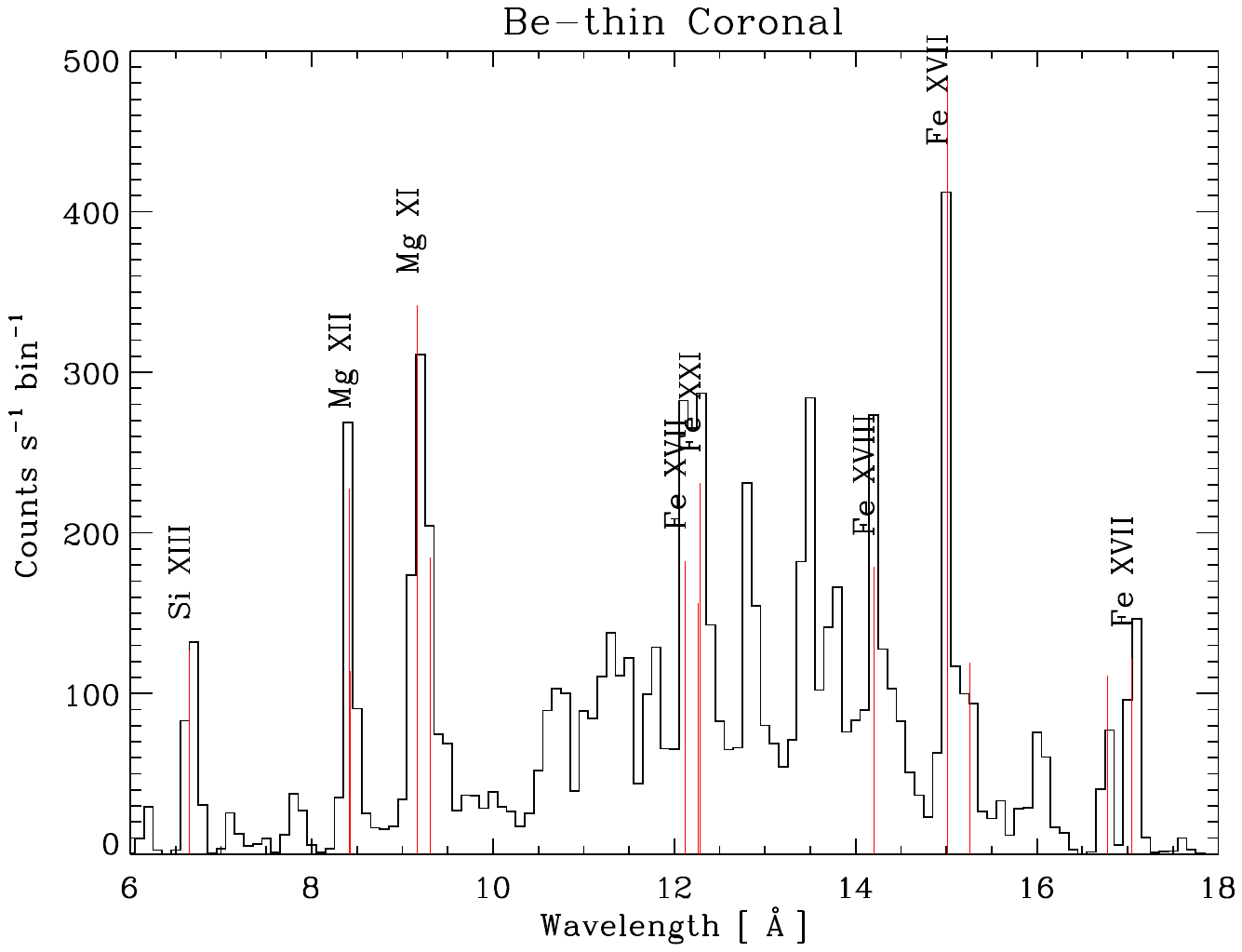}
}
\caption{Left panel: Simulated spectra from CHIANTI corresponding to the Be-thin filter of XRT using photospheric abundances. Right panel: Simulated spectra from CHIANTI corresponding to the Be-thin filter of XRT using coronal abundances. }
\label{fig:be_thin_phot}
\end{figure*}
\end{center}

Fig. \ref{fig:be_med_phot} shows the simulated spectra from CHIANTI, using the EM values derived from SDO/AIA, for the Be-med filter of XRT using photospheric \citep{2009ARA&A..47..481A} and coronal \citep{1992PhyS...46..202F} abundances in the left and right panel respectively. These coronal abundances are the ratio of high-FIP and low-FIP elements which is 3 times higher than the photospheric abundances. The plots show the dominance of low-FIP elements like Si and Mg lines for both the abundances. Additionally, the simulated spectra with the photospheric abundance indicates a contribution from high-FIP element Ne X while the one with the coronal abundance shows more contributions from low-FIP Fe lines. Similarly, Fig. \ref{fig:be_thin_phot} shows the simulated spectra from CHIANTI, using the EM values derived from SDO/AIA, for the Be-thin filter using photospheric and coronal abundances in the left and right panels respectively. The major contribution is from Fe XVII along with other low-FIP elements like Mg and Si. On the other hand, Be-med filter with photospheric abundances shows significant contribution from high-FIP ions like Ne X resulting in different filter ratio values.

\begin{center}
\begin{figure*}
\hspace{1.0cm}
\includegraphics[width=1\linewidth,angle=180]{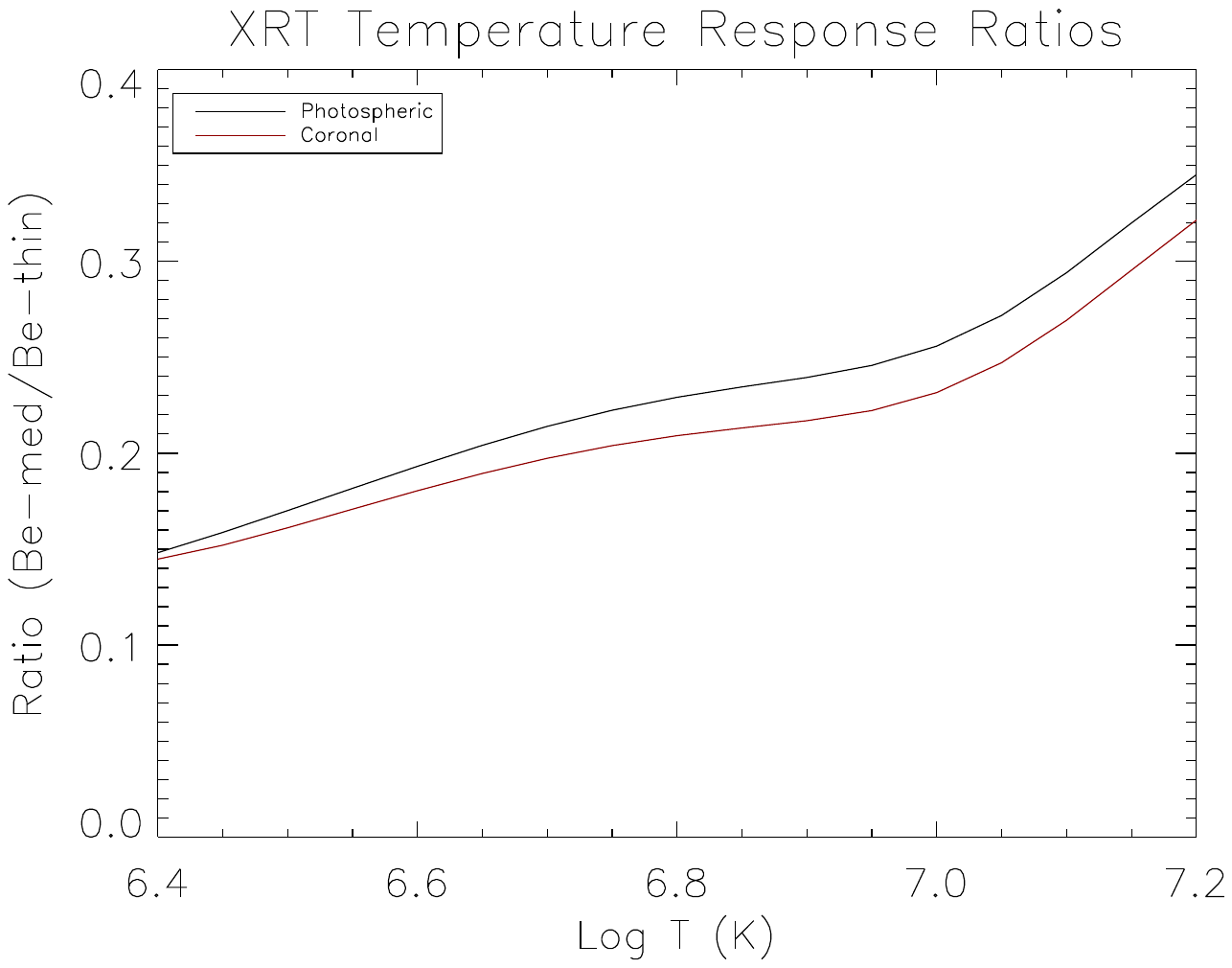}
\caption{The temperature response ratio for two filters of XRT (Be-med/Be-thin) w.r.t. the temperatures. The black line shows the ratios for photospheric abundance and the red line corresponds to coronal abundances.}
\label{fig:xrt_temp_ratio}
\end{figure*}
\end{center}

Fig. \ref{fig:xrt_temp_ratio} shows the temperature response ratio for the two filters of XRT (Be-med/Be-thin) w.r.t. the temperatures. The black line shows the ratios for photospheric abundance and the red line corresponds to coronal abundances. These show significant differences, resulting in the difference of effective temperatures calculated from XRT data using the Be-thin/Be-med filter ratio for photospheric and coronal abundances.

\end{document}